\documentclass[11pt]{article}
\usepackage{amsfonts,amsmath,amssymb}
\usepackage{enumerate}
\usepackage{hyperref}
\usepackage{bbm}
\usepackage{nicefrac}
\usepackage[all]{xy}
\usepackage{graphicx}
\usepackage{bm}

\usepackage{upgreek}

\usepackage{booktabs}

\newcommand{\be}{\begin{equation}}
\newcommand{\ee}{\end{equation}}

\newcommand{\bea}{\begin{eqnarray}}
\newcommand{\eea}{\end{eqnarray}}

\newcommand{\bes}{\begin{subequations}}
\newcommand{\ees}{\end{subequations}}

\newcommand{\cN}{{\cal N}}

\newcommand{\cH}{{\cal H}}

\def\sst#1{{\scriptscriptstyle #1}}

\def\0{{\sst{(0)}}}
\def\1{{\sst{(1)}}}
\def\2{{\sst{(2)}}}
\def\3{{\sst{(3)}}}
\def\4{{\sst{(4)}}}
\def\5{{\sst{(5)}}}
\def\6{{\sst{(6)}}}
\def\7{{\sst{(7)}}}
\def\8{{\sst{(8)}}}

\usepackage{multirow}
\usepackage{rotating}

\allowdisplaybreaks

\begin{document}

\makeatletter
\renewcommand{\theequation}{\thesection.\arabic{equation}}
\@addtoreset{equation}{section}
\makeatother

\begin{titlepage}

\begin{flushright}
\end{flushright}

\vspace{25pt}

   \begin{center}
   \baselineskip=16pt
   \begin{Large}\textbf{
AdS$_4$ solutions of massive IIA \\[12pt] from dyonic ISO(7) supergravity}
   \end{Large}

\vspace{40pt}

{Oscar Varela}

\vspace{25pt}

	\begin{small}
         
         {\it Center for the Fundamental Laws of Nature,\\
	Harvard University, Cambridge, MA 02138, USA } \\
	%ovarela@physics.harvard.edu

	\end{small}

\vskip 50pt

\end{center}

\begin{center}
\textbf{Abstract}
\end{center}

\begin{quote}

Explicit formulae are given for the consistent truncation of massive type IIA supergravity on the six-sphere to the SU(3)--invariant sector of $D=4$ $\cN=8$ supergravity with dyonic ISO(7) gauging. These formulae are then used to construct AdS$_4$ solutions of massive type IIA via uplift on $S^6$ of the critical points of the $D=4$ supergravity with at least SU(3) symmetry. We find a new $\cN=1$ solution with SU(3) symmetry, a new non-supersymmetric solution with SO(6) symmetry, and recover previously known solutions. We quantise the fluxes, calculate the gravitational free energies of the solutions and discuss the stability of the non-supersymmetric ones. Among these, a (previously known) G$_2$--invariant  solution is found to be stable.

\end{quote}

\vfill

\end{titlepage}

\tableofcontents

%%%%%%%%%%%%%%%%%%%%%%%%%%%%%%%%%%%%%%%%%%%%%%%%%%%%%%%%%%%%%%%%%%%%%%%%%%

\section{Introduction}

Supersymmetric solutions of string or M-theory containing four- or five-dimensional, for definiteness, anti-de-sitter (AdS) factors provide consistent backgrounds where the anti-de-Sitter/conformal field theory (AdS/CFT) correspondence \cite{Maldacena:1997re,Gubser:1998bc,Witten:1998qj}  can be formulated precisely. Well known early examples include the maximally supersymmetric Freund-Rubin direct product solutions $\textrm{AdS}_4 \times S^7$ \cite{Freund:1980xh} and $\textrm{AdS}_5 \times S^5$ \cite{Schwarz:1983qr} of $D=11$ and type IIB supergravity, respectively, where the spheres are equipped with their usual homogeneous round metrics with SO(8) and SO(6) symmetry. Less supersymmetric solutions in the Freund-Rubin class are obtained by either quotienting $S^7$ and $S^5$, or by altogether replacing them with more general classes of manifolds. For instance, Sasaki-Einstein seven- and five-manifolds (of which $S^7$ and $S^5$ are special examples) generically admit two Killing spinors and thus give rise to $\cN=2$ AdS$_4$ and AdS$_5$ solutions. Besides the finite list of homogeneous Freund-Rubin solutions (see \cite{Castellani:1983yg} for AdS$_4$ in $D=11$), infinite families of inhomogeneous spaces, for example in the Sasaki-Einstein class \cite{Gauntlett:2004hh, Gauntlett:2004yd}, give rise to infinite families of supersymmetric inhomogeneous Freund-Rubin solutions in $D=11$ and IIB.

Warped product solutions, both supersymmetric and non-supersymmetric, involving AdS$_4$ and AdS$_5$ in $D=11$ and type IIB, respectively, are also known. Smooth solutions of this type have been obtained by exploiting the consistent truncation of $D=11$ on $S^7$ \cite{deWit:1986iy,deWit:2013ija} and type IIB on $S^5$ (see \cite{Lee:2014mla,Ciceri:2014wya,Baguet:2015sma}) down to $\cN=8$ supergravity in $D=4$ with SO(8) \cite{deWit:1982ig} and $D=5$ with SO(6) \cite{Gunaydin:1984qu} gauging. Any solution of these gauged supergravities uplifts on the corresponding sphere to $D=11$ or type IIB. In particular, the (AdS) critical points of their scalar potentials respectively give rise to AdS$_4$ and AdS$_5$ solutions in $D=11$ and IIB. Some examples include \cite{deWit:1984va,Corrado:2001nv,Godazgar:2014eza} in $D=11$ and \cite{Pilch:2000ej} in type IIB. Together with the warp factor, these solutions are typically supported by internal values of the supergravity forms. The metrics on the internal $S^7$ and $S^5$ are usually inhomogeneous and display isometry groups smaller than SO(8) and SO(6). A two-step variant of this uplifting method was introduced in \cite{Maldacena:2000mw}, where a spherical truncation to some intermediate dimension, {\it e.g.} $D=7$ \cite{Nastase:1999cb,Nastase:1999kf}, followed by a further reduction on a suitable (usually hyperbolic) space was performed to obtain supersymmetric AdS$_4$ or AdS$_5$ solutions.

All of the above AdS$_4$ and AdS$_5$ direct product solutions and most (but not all, see {\it e.g.} \cite{Gauntlett:2000ng}) of the warped products are partially supported by a non-vanishing (`electric') Freund-Rubin-like term $\hat F_\4 \sim \textrm{vol(AdS)} _4$ in $D=11$ and  $\hat F_\5 \sim \textrm{vol(AdS)}_5$ in type IIB. This reflects that the corresponding solutions describe conformal phases of the M2, D3 and, for \cite{Gauntlett:2000ng}, M5 brane field theories. Since type IIA supergravity also contains a four-form field strength $\hat F_4$, it would be natural to expect that the above plethora of AdS$_4$ solutions in $D=11$ had a counterpart in type IIA. However, this is not the case: excluding the massless IIA solutions obtained by circle reduction from $D=11$, only a handful of AdS$_4$ solutions in type IIA are known, either smooth and sourcelss or singular and with sources. And most of them are only known numerically. 

Massive type IIA supergravity \cite{Romans:1985tz} does admit a direct product Freund-Rubin solution $\textrm{AdS}_4 \times S^6$ \cite{Romans:1985tz}, where $S^6$ is equipped with the usual, round, homogeneous, SO(7)-symmetric metric: see equation (\ref{SO7SolIIA}). However, unlike its maximally supersymmetric counterparts in $D=11$ \cite{Freund:1980xh}  and type IIB \cite{Schwarz:1983qr}, it breaks all supersymmetries. Furthermore, as it will be argued below, this solution is unstable. $G$-structure methods have been used to classify supersymmetric AdS$_4$ solutions in type IIA (as well as in other contexts) and, in some cases, these results have led to explicit classes of solutions. $\cN=1$ AdS$_4$ solutions with SU(3)-structure were classified in \cite{Behrndt:2004km} and \cite{Lust:2004ig}. An explicit class of $\cN=1$ solutions of massive type IIA involving the direct product of AdS$_4$ with a nearly-K\"ahler six-dimensional manifold was discovered in  \cite{Behrndt:2004km}. This class includes, in particular, the homogeneous $\cN=1$ direct product $\textrm{AdS}_4 \times S^6$, where the six-sphere is now regarded as $S^6 = \textrm{G}_2/\textrm{SU}(3)$ and the internal supergravity forms take values along the G$_2$--invariant nearly-K\"ahler forms. In stark contrast with the Sasaki-Einstein situation \cite{Gauntlett:2004hh, Gauntlett:2004yd}, the only analytically known nearly-K\"ahler six-manifolds are the four (and only four \cite{Butruille}) homogeneous cases\footnote{Only very recently, existence results of inhomogeneous nearly-K\"ahler metrics on $S^6$ and other manifolds have been given \cite{Foscolo:2015vqa}. See also \cite{Podesta09,Podesta:2012zz}. Analytic, closed form expressions for such metrics are not known.}, including $S^6 = \textrm{G}_2/\textrm{SU}(3)$. A `half-flat' generalisation of the nearly-K\"ahler condition \cite{Behrndt:2004km} for $\cN=1$ AdS$_4$ solutions with SU(3)-structure is known \cite{Lust:2004ig}. Also in that larger class, however, the only known solutions are homogeneous \cite{Koerber:2008rx} (see also \cite{Tomasiello:2007eq}).

Classifications of $\cN=1$ and $\cN=2$ AdS$_4$ solutions of type IIA with local SU(2) (that is, global $\textrm{SU}(3) \times \textrm{SU}(3)$) structure have also been carried out \cite{Grana:2005sn,Lust:2009zb,Lust:2009mb,Lust:2008zd}. However, these analyses lead to non-linear systems of first-order coupled differential equations for the supergravity fields that usually are very difficult to solve analytically. Some cohomogeneity-one $\cN=2$ AdS$_4$ solutions in this class are known numerically \cite{Petrini:2009ur,Aharony:2010af}. Only very recently, a combination of $G$-structure methods \cite{Grana:2005sn} with a two-step reduction {\it \`a la} Maldacena-N\'u\~nez \cite{Maldacena:2000mw} has produced explicit, analytic, $\cN=1$ warped product AdS$_4$ solutions in massive IIA with (typically, but not necessarily, hyperbolic) internal spaces with $\textrm{SU}(3) \times \textrm{SU}(3)$-structure \cite{Rota:2015aoa} (see also \cite{Apruzzi:2015wna}). Duality methods have been also used recently to obtain massless type IIA AdS$_4$ solutions with no Freund-Rubin term \cite{Lozano:2015cra}, similarly to \cite{Gauntlett:2000ng}. See \cite{Beck:2015hpa} for further recent classifications of supersymmetric AdS backgrounds of type IIA.

\begin{table}[t!]
\begin{center}
\renewcommand{\arraystretch}{1.5}
\scalebox{1.0}{
\begin{tabular}{cc|cccccc}
\noalign{\hrule height 1pt}
$\mathcal{N}$ & $G_{0}$ &  \phantom{abc}  & Stable? &  \phantom{abc} & $F$ &  \phantom{abc} & ref.  \\
\noalign{\hrule height 1pt}
$\mathcal{N}=2$ & $\textrm{SU}(3) \times \textrm{U}(1)$ && yes && $ 2^{1/3} \, 3^{1/6} \, 5^{-1} \, \pi \,  N^{5/3} k^{1/3}  $  && \cite{Guarino:2015jca}  \\
$\mathcal{N}=1$ & $\textrm{G}_{2}$ && yes & & $  2^{-7} \, 3^{7/6} \, 5^{3/2} \, \pi \,  N^{5/3} k^{1/3}  $ & & \cite{Behrndt:2004km} \\
$\mathcal{N}=1$ & $\textrm{SU}(3)$ && yes && $ 2^{-17/3} \, 3^{1/6} \, 5^{3/2} \, \pi \,  N^{5/3} k^{1/3}  $ && (\ref{SU3IIASolutionRescaled})
  \\
\hline
$\mathcal{N}=0$ & $\textrm{SO}(7)_+$ && no && $ 2^{7/3} \, 3^{2/3} \, 5^{-13/6} \, \pi \,  N^{5/3} k^{1/3}  $ && \cite{Romans:1985tz}  \\
$\mathcal{N}=0$ & $\textrm{SO}(6)_+$ && no && $ 2^{-1/2} \, 3^{2/3} \, 5^{-1} \, \pi \,  N^{5/3} k^{1/3}  $  && (\ref{SO6SolIIARescaled})  \\
$\mathcal{N}=0$ & $\textrm{G}_{2}$ && yes && $ 2^{-3} \, 3^{13/6} \, 5^{-1}  \,  \pi \,  N^{5/3} k^{1/3}  $ &&  \cite{Lust:2008zd} \\
\noalign{\hrule height 1pt}
\end{tabular}
}
\caption{Summary of AdS$_4$ solutions of massive type IIA supergravity with supersymmetry $\cN$ and bosonic symmetry $G_0$ that are obtained from uplift of the analytical critical points of $D=4$ $\cN=8$ dyonic ISO(7) supergravity with at least SU(3) symmetry. The stability against perturbations contained in the $D=4$ supergravity, the gravitational free energy $F$ of the solutions and the reference where they were first found are also shown.}
\label{Table:SU3Points}
\end{center}
\end{table}

In this paper, we obtain new supersymmetric and non-supersymmeric AdS$_4$ solutions of massive type IIA and recover some previously known ones. We do this by-passing altogether the integration of difficult non-linear  systems of first \cite{Grana:2005sn,Lust:2009zb,Lust:2009mb,Lust:2008zd} or second order (see appendix \ref{SU3InvAdS4Sols}) differential equations. Instead, we use consistent truncation methods. In \cite{Guarino:2015jca,Guarino:2015vca} it was found that massive type IIA supergravity admits a consistent truncation on $S^6$ down to $D=4$ $\cN=8$ supergravity with gauge group ISO(7)  \cite{Guarino:2015qaa}. The gauging is of the dyonic type discussed in \cite{Dall'Agata:2012bb,Dall'Agata:2014ita}. By the consistency of the truncation, all $D=4$ solutions uplift to massive IIA. In particular, the (AdS) critical points of the scalar potential give rise to AdS$_4$ solutions of massive type IIA, preserving supersymmetry in the process if present. This is exactly as in the uplifts to $D=11$ and type IIB mentioned above. 

Specifically, we uplift the critical points of the $D=4$ supergravity with at least SU(3) symmetry that were classified in \cite{Guarino:2015qaa} (see table 3 of that reference for a summary) to obtain smooth, supersymmetric and non-supersymmetric, direct or warped product solutions of AdS$_4$ and $S^6$. All the solutions are supported by a Freund-Rubin electric flux $\hat F_\4 \sim \textrm{vol(AdS)} _4$ (or equivalently, by $\hat F_6$ flux on $S^6$) and some of them by a warp factor and internal supergravity forms as well. For each solution, the metric on $S^6$ displays an isometry related  to the symmetry of the $D=4$ critical point it uplifts from. In particular, we obtain new solutions with $\cN=1$ and SU(3) symmetry and $\cN=0$ and SO(6)$_+$ symmetry, the subscript indicating that the corresponding $D=4$ point is supported by $D=4$ scalars, rather than pseudoscalars. We comment on generalisations and on the stability of the non-supersymmetric solutions. We also discuss flux quantisation and compute the gravitational free energy of the solutions. See table \ref{Table:SU3Points} for a summary.

We obtain these solutions by first working out explicit consistent truncation formulae of massive type IIA on $S^6$ to the full, dynamical, SU(3)--invariant sector of dyonic ISO(7) supergravity. We do this by particularising the general $\cN=8$ formulae given in \cite{Guarino:2015jca,Guarino:2015vca}. In other words, we explicitly obtain the full non-linear embedding of the entire, dynamical SU(3)--invariant sector of dyonic ISO(7) supergravity into massive type IIA at the level of the metric, dilaton and IIA form potentials (see equation (\ref{KKSU3sectorinIIA})). These formulae allow for the uplift of any solution of the $D=4$ theory, not only critical points of the potential. As emphasised in \cite{Guarino:2015vca}, these formulae do not depend on the Romans mass, which only enters the IIA embedding of the $D=4$ bosons through the field strengths and scalar covariant derivatives. These formulae are thus also valid to uplift solutions of the SU(3)--invariant sector of the purely electric ISO(7) gauging \cite{Hull:1984yy} to massless type IIA. These should necessarily have non-constant scalars or non-trivial profiles for the vector fields, given the absence of critical points for the electric ISO(7) gauging.

Section \ref{sec:SU3sectorinIIA} discusses the embedding of the entire SU(3) sector of ISO(7) supergravity into type IIA. We discuss the regularity, symmetry and supersymmetry of the embedding and, in section \ref{sec:urthertruncs}, particularise it to some subsectors. Section  \ref{sec:SU3sectorinIIA} also contains the massive IIA field strengths evaluated on constant scalar configurations. Particularising further to the individual critical points in the SU(3) sector, we obtain the solutions contained\footnote{Two of the previously known solutions we recover have also been rederived in the recent \cite{Pang:2015vna} using the uplifting formulae of \cite{Guarino:2015jca,Guarino:2015vca}.} in section \ref{sec:AdSsolutions}. The paper concludes with two technical appendices. The foliation of $S^6$ with $S^5$ leaves, where $S^5$ is equipped with its usual Sasaki-Einstein structure, is described in appendix \ref{subset:SE5S6}. This is the structure that naturally emerges in the embedding studied in section \ref{sec:SU3sectorinIIA} and the solutions of section \ref{sec:AdSsolutions}. Appendix \ref{SU3InvAdS4Sols} specifies the system of equations, derived from the massive type IIA field equations, that the solutions of section  \ref{sec:AdSsolutions} must obey. 

%%%%%%%%%%%%%%%
%%%%%%%%%%%%%%%
\section{\mbox{Truncation to the SU(3) sector of ISO(7) supergravity}} \label{sec:SU3sectorinIIA}
%%%%%%%%%%%%%%%
%%%%%%%%%%%%%%%

We will now particularise the general $\cN=8$ consistent embedding formulae of  \cite{Guarino:2015jca,Guarino:2015vca} to the SU(3)--invariant bosonic sector of dyonic ISO(7) supergravity. The embedding formulae are naturally expressed, at the level of the IIA metric, dilaton and form potentials, in terms of $D=4$ SU(3)-singlet fields in the restricted duality hierarchy introduced in \cite{Guarino:2015qaa}. The relevant $D=4$ field content includes
\begin{eqnarray} \label{fieldContentHierarchy}
\textrm{1 metric} & : & \quad ds_4^2 \; ,  \nonumber \\
\textrm{6 scalars} & : & \quad \varphi \; , \;  \chi \; , \; \phi \; , \;  a\; , \;  \zeta \; , \;  \tilde \zeta \; , \nonumber \\
\textrm{2 electric vectors and their magnetic duals} & : & \quad A^0 \; , \;  A^1 \; , \; \tilde{A}_0 \; , \;  \tilde{A}_1 \; , \nonumber \\
\textrm{3 two-forms} & : & \quad B^0 \; , \;  B_1 \; , \; B_2 \; , 
\nonumber \\
\textrm{2 three-forms} & : & \quad C^0 \; , \;  C^1 \; ,
\end{eqnarray}
see section 3 of \cite{Guarino:2015qaa}. All these fields are real. The scalars parametrise a submanifold
\begin{eqnarray} 
\label{ScalManN=2}
\frac{\textrm{SU}(1,1)}{\textrm{U}(1)} \times  \frac{\textrm{SU}(2,1)}{\textrm{SU}(2) \times \textrm{U}(1)} \,   
\end{eqnarray}
of E$_{7(7)}/$SU(8). As discussed in \cite{Guarino:2015qaa} following \cite{Bergshoeff:2009ph}, not all of the fields in (\ref{fieldContentHierarchy}) carry independent degrees of freedom. The four-form field strengths of the three-form potentials $C^0$, $ C^1$ can be dualised into functions on the scalar manifold (\ref{ScalManN=2}), etc. See (3.9), (3.13)--(3.15) of \cite{Guarino:2015qaa} for the definitions of the field strengths of the form potentials in (\ref{fieldContentHierarchy}) and (3.17)--(3.19) of \cite{Guarino:2015qaa} for the duality conditions. These dualisations can be used to write the consistent embedding into the type IIA field strengths in terms of independent degrees of freedom only.

In this section we give the full non-linear embedding in type IIA of the SU(3) sector of ISO(7) supergravity, at the level of the IIA metric, dilaton and form potentials, using the restricted tensor hierarchy (\ref{fieldContentHierarchy}). We also give the field strengths evaluated on constant scalar configurations, and employ the duality relations for the four-form field strengths to express the Freund-Rubin term as a function of the $D=4$ scalars (in full generality, not only for constant scalars). We also discuss the regularity, symmetry and supersymmetry of the embedding. Further examples on the use of the dualisation conditions can be found in section \ref{sec:urthertruncs}. See appendix A of \cite{Guarino:2015vca} for our type IIA conventions.

%%%%%%%%%%%%%%%
%%%%%%%%%%%%%%%
\subsection{Consistent embedding formulae}
\label{subsec:SU3UpliftSubsec}

%%%%%%%%%%%%%%%
%%%%%%%%%%%%%%%

The embedding of the SU(3)--invariant $D=4$ $p$-forms, $p=1,2,3$, in (\ref{fieldContentHierarchy}) into the $\cN=8$ SL(7)-covariant restricted tensor hierarchy \cite{Guarino:2015qaa} of the ISO(7) gauging was given in equation (3.4) of that reference. We can bring those definitions to the uplifting formulae (3.12), (3.13) of \cite{Guarino:2015vca}, and use the $S^6$ conventions of appendix \ref{subset:SE5S6}. Then, some simple algebra allows us to obtain the ten-dimensional embedding of these $D=4$ $p$-forms. In contrast, the IIA embedding of the $D=4$ scalars is calculationally more intense. This can be achieved by bringing the SU(3)--invariant $D=4$ scalar matrix, given in appendix D.1 of \cite{Guarino:2015qaa}, to the uplifting formulae (10) of \cite{Guarino:2015jca} (or, equivalently, (3.14)--(3.18) of \cite{Guarino:2015vca}). The non-vanishing components of the $D=4$ scalar matrix occur along the invariant metric $\delta_{ij}$, $i=1, \ldots , 6$, and tensors $J$, $\Omega$ that define the SU(3)--holonomy of the $\mathbb{R}^6$ factor in the ambient $\mathbb{R}^7 = \mathbb{R}^6  \times \mathbb{R}$ where $S^6$ is embedded. The embedding coordinates $\mu^I$, $I=1 , \ldots, 7$, adapted to this setting are given in  (\ref{eq:SplitS6intoS5}), and the Killing vectors and their derivatives in (\ref{KillingFoliation}) and (\ref{KillingDerFoliation}). Using these formulae, we find that all contractions of $\mu^I$ and $\partial_m \mu^I$ with the scalar matrix happen to occur through the combinations (\ref{SE5intermsofmu}) that define the Sasaki-Einstein structure $\bm{J}, \bm{\Omega}, \bm{\eta}$ of the $S^5$ within $S^6$.

In order to present the result, it is useful to define the following combinations of $D=4$ scalars
\begin{eqnarray} \label{eq:XY}
X \equiv 1 + e^{2\varphi} \chi^2  \; , \qquad 
Y \equiv 1 +\tfrac14 e^{2\phi}  (\zeta^2 + \tilde\zeta^2 )   \; , 
\end{eqnarray}
and $D=4$ scalars together with an angle $\alpha$ on $S^6$ (see appendix \ref{subset:SE5S6}), 
\begin{eqnarray} \label{Delta1}
&& \Delta_1 = \big( e^{\varphi} +\tfrac14 e^{2\phi + \varphi}  (\zeta^2 + \tilde\zeta^2 ) \big) \sin^2 \alpha  + e^{2\phi-\varphi} \big( 1+ e^{2\varphi} \chi^2 \big) \cos^2 \alpha \; , \\[4pt]
\label{Delta2}
&& \Delta_2 = e^{\varphi}  \sin^2 \alpha  + e^{2\phi-\varphi}   \cos^2 \alpha \; .
\end{eqnarray}
With these definitions, the complete nonlinear embedding of the SU(3)--invariant field content (\ref{fieldContentHierarchy}) of ISO(7) supergravity into type IIA reads, in ten-dimensional Einstein frame,
{\setlength\arraycolsep{0pt}
\begin{eqnarray} \label{KKSU3sectorinIIA}
&& d\hat{s}_{10}^2 =  e^{\frac18 (2\phi-\varphi)} X^{1/4}  \Delta_1^{1/2}  \Delta_2^{1/8}  \Big[ \; ds_4^2   \nonumber \\
&& \qquad \qquad \qquad \qquad + g^{-2} e^{-2\phi+\varphi}   X^{-1}  d\alpha^2  +  g^{-2}  \sin^2 \alpha  \Big( \Delta_1^{-1} ds^2 ( \mathbb{CP}^2 ) + X^{-1} \Delta_2^{-1} \bm{\tilde{\eta}}^2  \Big) \Big] \; ,  \nonumber \\[12pt]
&& e^{\hat \phi}  =  e^{\frac14 (6\phi +5\varphi)} X^{-1/2}  \Delta_1^{-1}  \Delta_2^{3/4} \; , \nonumber \\[12pt]
&& \hat  A_\3 = \cos^2 \alpha \, \big( C^0 + A^0 \wedge B^0 \big) + \sin^2 \alpha \, C^1   \nonumber \\
&& \qquad -g^{-1} \sin \alpha \cos \alpha \, \Big( B_1 + \tfrac12 A^0 \wedge \tilde{A}_0 + \tfrac16 A^1 \wedge \tilde{A}_1 \Big)  \wedge d \alpha - \tfrac13 \, g^{-1} \, \sin^2\alpha \ B_2 \wedge \bm{\tilde{\eta}}  \nonumber \\
&& \qquad + \tfrac13 \, g^{-2} \,  \tilde{A}_1\wedge  \Big[  \sin^2\alpha \,  \bm{J}  + \sin\alpha \cos \alpha \,  d\alpha \wedge \bm{\tilde{\eta}}  \Big]  \nonumber \\
&& \qquad -\cos \alpha \, A^0 \wedge   \Big[ g^{-2} e^{2\varphi} \,  \chi \,  X^{-1}   \sin \alpha  \,  d\alpha \wedge \bm{\tilde{\eta}}    \nonumber \\
&& \qquad +   g^{-2} e^{2\phi + \varphi} \,  \chi \,  \Delta_1^{-1}   \sin^2 \alpha \cos \alpha \, \bm{J}  +\tfrac12 g^{-2} e^{2\phi +\varphi}  \Delta_1^{-1} \sin^3 \alpha \,  \big( \tilde \zeta \, \textrm{Re} \, \bm{\Omega} - \zeta \, \textrm{Im} \, \bm{\Omega} \big) \Big] \nonumber \\
&&  \qquad -  g^{-3} e^{2\phi + \varphi} a \, \chi \,  \Delta_1^{-1}   \sin^3 \alpha \cos \alpha \, \bm{J} \wedge d\alpha -  g^{-3} e^{ \varphi} \chi \, Y \, \Delta_1^{-1}   \sin^4 \alpha  \, \bm{J} \wedge \bm{\tilde{\eta}} \nonumber \\
&& \qquad + \tfrac12 g^{-3}  \sin^2 \alpha \left(  \big( \zeta \,  \textrm{Re} \, \bm{\Omega} +  \tilde \zeta \,  \textrm{Im} \, \bm{\Omega}  \big) 
 - e^{2\phi +\varphi} a \,  \Delta_1^{-1} \sin^2 \alpha  \big( \tilde \zeta \,  \textrm{Re} \, \bm{\Omega} -  \zeta \,  \textrm{Im} \, \bm{\Omega}  \big)  \right) \wedge d\alpha \nonumber \\
&& \qquad + \tfrac12 g^{-3} e^{2\phi - \varphi} X  \, \Delta_1^{-1} \sin^3 \alpha \cos \alpha  \Big( \tilde \zeta \,  \textrm{Re} \, \bm{\Omega} -  \zeta \,  \textrm{Im} \, \bm{\Omega}  \Big) \wedge \bm{\tilde{\eta}}  \; , \nonumber \\[12pt]
&& \hat  B_\2 = -\cos \alpha \, B^0 + g^{-1} \sin \alpha \, \tilde{A}_0 \wedge d\alpha +g^{-2} e^{2\varphi} \,  \chi \,  X^{-1}   \sin \alpha  \,  d\alpha \wedge \bm{\tilde{\eta}}    \nonumber \\
&& \qquad +   g^{-2} e^{2\phi + \varphi} \,  \chi \,  \Delta_1^{-1}   \sin^2 \alpha \cos \alpha \, \bm{J}  +\tfrac12 g^{-2} e^{2\phi +\varphi}  \Delta_1^{-1} \sin^3 \alpha \,  \big( \tilde \zeta \, \textrm{Re} \, \bm{\Omega} - \zeta \, \textrm{Im} \, \bm{\Omega} \big) \; , \nonumber \\[12pt]
&& \hat  A_\1 = -\cos \alpha \, A^0 - g^{-1} a \sin \alpha \, d \alpha + g^{-1} e^{-\varphi} \big( X-Y  \big) \Delta_2^{-1} \sin^2 \alpha \cos \alpha \, \bm{\tilde{\eta}} \; .
\end{eqnarray}
}The metric $ds^2 ( \mathbb{CP}^2 ) $ is the usual  Fubini-Study metric on the $\mathbb{CP}^2$ base of the $S^5$ within $S^6$, normalised so that the Ricci tensor is six times the metric, and 
\begin{equation}
\bm{\tilde{\eta}} \equiv \bm{\eta} + g\, A^1 \equiv  d\psi + \sigma + g\, A^1 \; .
\end{equation}
Here, $\psi$ is the angle on the Hopf fiber of $S^5$ and  $\sigma$ is a one-form potential for $\bm{J}$: $ d\sigma =2 \, \bm{J}$. These embedding formulae depend on the (non-vanishing) $D=4$ electric gauge coupling $g$, but not on the magnetic coupling $m$. Thus, they simultaneously describe the embedding of the SU(3)--invariant sector of the purely electric, $m = 0$, and dyonic, $m \neq 0$, ISO(7) gauging into massless and massive, respectively, type IIA supergravity. 

The field strengths corresponding to the form potentials in (\ref{KKSU3sectorinIIA}) can be computed using their definitions, see (A.4) of \cite{Guarino:2015vca} for our conventions. For simplicity, here we present them for constant $D=4$ scalars and vanishing $D=4$ two- and three-form field strength contributions. The Romans mass is given by \cite{Guarino:2015jca} 
\begin{equation}
\label{Romans=m}
\hat F_\0 = m \; , 
\end{equation}
and $\hat F_\4$, $\hat H_\3$ and $\hat F_\2$ by

\newpage 
{\setlength\arraycolsep{0pt}
\begin{eqnarray} \label{KKSU3FieldStrengths}
&& \hat  F_\4 = U \,  \textrm{vol}_4  \nonumber\\
&& \qquad + \Big( m g^{-4} e^{4\phi + 2\varphi} \Delta_1^{-2}  \big[  \tfrac14 (\zeta^2 + \tilde  \zeta^2) \sin^2 \alpha + \chi^2 \cos^2\alpha   \big] -4 g^{-3} e^{\varphi} \chi \, Y  \, \Delta_1^{-1}   \Big) \sin^4 \alpha \,  \textrm{vol} (\mathbb{CP}_2)  \nonumber \\
&& \qquad  + \Big[ mg^{-4} e^{2\phi + 3\varphi} \chi^2 \, X^{-1} \, \Delta_1^{-1}  \nonumber \\
&& \qquad\qquad -  g^{-3} e^{ \varphi} \chi \,  \Delta_1^{-2} \Delta_2^{-1}    \Big( e^\varphi X^{-1} Y  \big(  e^{2\phi-\varphi} X (X+Y) + 2 e^\varphi Y^2  \big) \sin^4 \alpha  \nonumber \\
&& \qquad\qquad\qquad\qquad\qquad\qquad\;  +   \big(  e^{4\phi-2\varphi} X (X+Y) + 6 e^{2\phi} Y^2  \big)  \sin^2 \alpha   \cos^2 \alpha \nonumber \\
&& \qquad\qquad\qquad\qquad\qquad\qquad\; + 4e^{4\phi-2\varphi} XY  \cos^4 \alpha  \Big) \Big] \sin^3 \alpha \cos\alpha \, \bm{J} \wedge d\alpha \wedge \bm{\eta}    \nonumber \\
&& \qquad + \tfrac12 \Big[ m g^{-4} e^{2\phi +3 \varphi} \chi \, X^{-1}  \, \Delta_1^{-1} 
 \nonumber \\
&& \qquad \qquad   -g^{-3}  \,  \Delta_1^{-2} \Delta_2^{-1}    \Big( e^\varphi  Y  \big(  e^{2\phi} X + 3 e^{2\varphi} Y \big) \sin^4 \alpha  \nonumber \\
&& \qquad\qquad   \quad\qquad\qquad\qquad\; +   \big(  e^{4\phi-\varphi} X (X-Y) + 2 e^{2\phi+\varphi} Y (Y+3X)  \big)  \sin^2 \alpha   \cos^2 \alpha \nonumber \\
&&  \qquad\qquad   \quad\qquad\qquad\qquad\;   +   \big(  e^{4\phi-\varphi} X (3X+2Y) - e^{6\phi-3\varphi} X^2  \big)  \cos^4 \alpha  \Big) \Big] \nonumber \\
&& \qquad \qquad  \qquad\qquad   \quad\qquad\qquad\qquad\;   \times  \sin^4 \alpha  \, \big( \tilde  \zeta \, \textrm{Re} \, \bm{\Omega} -  \zeta \, \textrm{Im} \, \bm{\Omega} \big)  \wedge d\alpha \wedge \bm{\eta}  \; , \nonumber \\[14pt]
&& \hat  H_\3 =   -g^{-2} e^{2\phi + \varphi} \,  \chi \,  \Delta_1^{-2}  \Big( e^\varphi X^{-1} Y  \big(  X + 2 e^{-2\phi+2\varphi} Y  \big) \sin^2 \alpha +  \big(  4e^\varphi Y - e^{2\phi-\varphi} X  \big)   \cos^2 \alpha\Big) \nonumber \\
&&  \qquad\qquad\qquad\qquad\qquad \times  \sin^3 \alpha  \, \bm{J} \wedge d\alpha   \nonumber \\
&& \qquad + \tfrac12 g^{-2} e^{2\phi +\varphi}  \Delta_1^{-2}  \Big(  \big(  e^\varphi \, Y + 2 e^{2\phi-\varphi} X  \big) \sin^2 \alpha + 3e^{2\phi-\varphi } X  \cos^2 \alpha\Big) \sin^2 \alpha \cos \alpha \nonumber \\
&& \qquad\qquad\qquad\qquad\qquad \times   \big( \tilde  \zeta \, \textrm{Re} \, \bm{\Omega} - \zeta \, \textrm{Im} \, \bm{\Omega} \big) \wedge d\alpha   \nonumber \\
&& \qquad -\tfrac32 g^{-2} e^{2\phi +\varphi}  \Delta_1^{-1} \sin^3 \alpha \,  \big(  \zeta \, \textrm{Re} \, \bm{\Omega} + \tilde \zeta \, \textrm{Im} \, \bm{\Omega} \big) \wedge \bm{\eta}  \; , \nonumber \\[14pt]
%
%
%
%
%%
%&& \hat  F_\2 = \Big( m g^{-2} e^{2\phi+\varphi} \chi \, \Delta_1^{-1} + 2 g^{-1} e^{-\varphi} \big( X-Y  \big) \Delta_2^{-1} \Big) \sin^2 \alpha \cos \alpha \,  \bm{J} \nonumber  \\
%%
%&& \qquad \quad +\tfrac12 m g^{-2} e^{2\phi +\varphi}  \Delta_1^{-1} \sin^3 \alpha \,  \big( \tilde \zeta \, \textrm{Re} \, \bm{\Omega} - \zeta \, \textrm{Im} \, \bm{\Omega} \big)
%\nonumber  \\
%%
%&& \qquad \quad + \Big[ m g^{-2} e^{2 \varphi} \chi \, X^{-1} + g^{-1} e^{-\varphi} \big( X-Y  \big) \Delta_2^{-2}  \nonumber  \\
%%
%&& \qquad \qquad \times \Big( 2 \, e^{2\phi-\varphi} \cos^4\alpha  + e^{2\phi-\varphi} \cos^2\alpha \sin^2\alpha  - e^{\varphi} \sin^4 \alpha  \Big) \Big]  d\alpha \wedge  \bm{\eta} \; .
%%
%%
%&& \hat  F_\2 = \Big( m g^{-2} e^{2\phi+\varphi} \chi \, \Delta_1^{-1} + 2 g^{-1} e^{-\varphi} \big( X-Y  \big) \Delta_2^{-1} \Big) \sin^2 \alpha \cos \alpha \,  \bm{J} \nonumber  \\
%%
%&& \qquad \quad +\tfrac12 m g^{-2} e^{2\phi +\varphi}  \Delta_1^{-1} \sin^3 \alpha \,  \big( \tilde \zeta \, \textrm{Re} \, \bm{\Omega} - \zeta \, \textrm{Im} \, \bm{\Omega} \big)
%  \\
%%
%&& \qquad \quad + \Big[ m g^{-2} e^{2 \varphi} \chi \, X^{-1}   \nonumber  \\
%%
%&& \qquad \qquad + g^{-1} e^{-\varphi} \big( X-Y  \big) \Delta_2^{-2} \Big( 2 \, e^{2\phi-\varphi} \cos^4\alpha  + e^{2\phi-\varphi} \cos^2\alpha \sin^2\alpha  - e^{\varphi} \sin^4 \alpha  \Big) \Big]  \sin\alpha \, d\alpha \wedge  \bm{\eta} \; .
%\nonumber 
%%
%
&& \hat  F_\2 = \Big( m g^{-2} e^{2\phi+\varphi} \chi \, \Delta_1^{-1} + 2 g^{-1} e^{-\varphi} \big( X-Y  \big) \Delta_2^{-1} \Big) \sin^2 \alpha \cos \alpha \,  \bm{J} \nonumber  \\
&& \qquad \quad +\tfrac12 m g^{-2} e^{2\phi +\varphi}  \Delta_1^{-1} \sin^3 \alpha \,  \big( \tilde \zeta \, \textrm{Re} \, \bm{\Omega} - \zeta \, \textrm{Im} \, \bm{\Omega} \big)
  \\
&& \qquad \quad + \Big[ m g^{-2} e^{2 \varphi} \chi \, X^{-1}  + g^{-1} e^{-\varphi} \big( X-Y  \big) \Delta_2^{-2} \Big( 2 \, e^{2\phi-\varphi} \cos^4\alpha  
 \nonumber  \\
&& \qquad \qquad \qquad \qquad \qquad \qquad \qquad \qquad \qquad \; \;  + e^{2\phi-\varphi} \cos^2\alpha \sin^2\alpha  - e^{\varphi} \sin^4 \alpha  \Big) \Big]  \sin\alpha \, d\alpha \wedge  \bm{\eta} \; .
\nonumber 
\end{eqnarray}
}

In contrast to the gauge potentials in (\ref{KKSU3sectorinIIA}), the field strengths (\ref{KKSU3FieldStrengths}) now do depend on the $D=4$ magnetic gauge coupling $m$. This is a generic feature: see \cite{Guarino:2015vca} for further discussion. By the consistency of the truncation, the metric and dilaton in (\ref{KKSU3sectorinIIA}) and the field strengths (\ref{KKSU3FieldStrengths}) solve all the field equations of massless or massive, if $m =0$ or $m \neq 0$, type IIA supergravity provided the $D=4$ field equations on constant scalar configurations are imposed. In other words, the field strengths (\ref{KKSU3FieldStrengths}) solve the type IIA field equations when evaluated on critical points of the $D=4$ scalar potential. Critical points in the ISO(7) gauging only exist for $m \neq 0$, {\it i.e.}~in the dyonic case. See section \ref{sec:AdSsolutions} for the resulting AdS$_4$ solutions of massive IIA. Let us emphasise that the configuration (\ref{KKSU3sectorinIIA}) is valid, more generally, for dynamical scalars and $p$-forms in the $D=4$ hierarchy (\ref{fieldContentHierarchy}). The corresponding field strengths (namely, the extension of (\ref{KKSU3FieldStrengths}) to include the derivatives of the scalars along with the field strengths of the $p$-forms in  (\ref{fieldContentHierarchy})) obey the type IIA field equations provided their $D=4$ counterparts are imposed.

Following the strategy of \cite{Guarino:2015vca}, the Freund-Rubin term in $\hat F_\4$ can be determined by using the dualisation conditions for the $D=4$ four-form field strengths. In the case at hand, the Freund-Rubin term $U \, \textrm{vol}_4 = \cH_\4^{IJ} \mu_I \mu_J$ \cite{Guarino:2015vca} reduces to
\begin{equation} \label{eq:FR1}
U \, \textrm{vol}_4 = H_\4^0 \cos^2 \alpha +H_\4^1 \sin^2 \alpha \; , 
\end{equation}
where  $H_\4^0$ and $H_\4^1 $ are the four-form field strengths of the three-form potentials in (\ref{fieldContentHierarchy}) above (see (3.15) of \cite{Guarino:2015qaa}). Using the dualisation relations given in (3.19) of \cite{Guarino:2015qaa}, the function $U$ in (\ref{eq:FR1}) and (\ref{KKSU3FieldStrengths}) becomes
{\setlength\arraycolsep{2pt}
\begin{eqnarray}
\label{USU3}
U &=& \Big[  \tfrac{1}{2} \, g \,  \big(1+e^{2\varphi} \chi^2 \big)  \Big( 12 \, e^{2 \phi - \varphi } - 2  \,  e^{4 \phi - 3\varphi }  \big(1+e^{2\varphi} \chi^2 \big)^2  - 3 e^{4 \phi +\varphi } \chi^2 \big( \zeta^2 +\tilde{\zeta}^2 \big) \Big) \nonumber \\[4pt]
&& \quad + m\, e^{4 \phi + 3\varphi } \chi^3 \Big] \cos^2\alpha  \nonumber \\[4pt]
&& +  \Big[ \tfrac{1}{2} \, g \Big( 8 \, e^{\varphi } + 2 \, e^{2 \phi - \varphi }  \big(1+e^{2\varphi} \chi^2 \big) + e^{2\phi +\varphi} \big( \zeta^2 + \tilde \zeta^2 \big)   \big(1-3 \, e^{2\varphi} \chi^2 \big)  \nonumber  \\[4pt]
&&  \qquad  -\tfrac12 \, e^{4\phi +\varphi} \big( \zeta^2 + \tilde \zeta^2 \big) \chi^2   \big(1+e^{2\varphi} \chi^2 \big)
 -  \tfrac14 \, e^{4\phi +\varphi} \big( \zeta^2 + \tilde \zeta^2 \big)^2   \big(1+3 \, e^{2\varphi} \chi^2 \big) \Big) \nonumber \\[4pt]
&& \quad + \tfrac14 \, m\, e^{4 \phi + 3\varphi } \chi \,   \big( \zeta^2 + \tilde \zeta^2 \big)  \Big]  \sin^2\alpha  \; .
\end{eqnarray}
}This expression can be equivalently obtained from the master formula (3.43) of \cite{Guarino:2015vca}. 

Although the field strengths (\ref{KKSU3FieldStrengths}) are evaluated for constant scalars, the Freund-Rubin term (\ref{USU3}) is general: it is valid for both constant and dynamical scalars. In general, the Freund-Rubin term can be expressed in terms of the scalar potential, the covariant derivatives of the three-form field strengths in the restricted $D=4$ tensor hierarchy, and the vector field strengths, see equation (3.36) of \cite{Guarino:2015vca}. Upon dualisation, the last two contributions become related to scalar equations of motion. In particular, the dualised scalar contributions should be related to derivatives of the scalar potential. This can be exhibited with the explicit parametrisation of the SU(3)-invariant sector that we are using. Indeed, the function $U$ in (\ref{USU3}) that defines the Freund-Rubin term can be checked to be related to the SU(3)-invariant $D=4$ scalar potential $V$ (given in (3.11) of \cite{Guarino:2015qaa}) and its derivatives:
{\setlength\arraycolsep{2pt}
\begin{eqnarray} \label{UintermsofV}
g \, U &=& -\tfrac{1}{3} \, V -\tfrac16 \Big(  \partial_\phi V -2 \, \partial_\varphi V +2 \chi \, \partial_\chi V - \zeta \, \partial_\zeta V - \tilde{\zeta} \,  \partial_{\tilde{\zeta}} V \Big) \, \cos^2\alpha    \nonumber \\
&& \qquad  \quad  +\tfrac{1}{12} \Big(  \partial_\phi V - \zeta \, \partial_\zeta V - \tilde{\zeta} \,  \partial_{\tilde{\zeta}} V \Big) \, \sin^2\alpha \; .
\end{eqnarray}
}
At a critical point of the scalar potential, this expression reduces to
{\setlength\arraycolsep{2pt}
\begin{eqnarray} \label{UintermsofVCritical}
g \, U |_0 &=& -\tfrac{1}{3} \, V_0 \; , 
\end{eqnarray}
}in agreement with the general $\cN=8$ discussion of \cite{Guarino:2015vca}. In (\ref{UintermsofVCritical}), $|_0$ and $V_0$ denote evaluation at a critical point. See \cite{Godazgar:2015qia,Varela:2015ywx} for a related discussion in a $D=11$ on $S^7$ context.

\subsection{Regularity, symmetry and supersymmetry} \label{sec:Regularity}

The local geometry corresponding to the type IIA configuration (\ref{KKSU3sectorinIIA}) can be regarded as a foliation, parametrised by $\alpha$, of $S^6$ with $S^5$ leaves. The $S^5$, in turn, comes naturally equipped with its Sasaki-Einstein structure. Namely, the $S^5$ can be viewed as a U(1), parametrised by $\psi$, fibered over $\mathbb{CP}^2$, with fibers stretched or squashed (depending on the values of the $D=4$ scalars) as a function of $\alpha$. It is easy to see that this geometry extends globally in a smooth fashion provided the angles $\alpha$, $\psi$ are given the ranges
\begin{eqnarray} \label{eq:ranges}
0 \leq \alpha \leq \pi \; , \qquad 
0 \leq \psi \leq 2 \pi \; . 
\end{eqnarray}
Globally, (\ref{KKSU3sectorinIIA}) is defined on a smooth $S^2$ bundle, parametrised by $\alpha, \psi$, over $\mathbb{CP}^2$. These angles do indeed parametrise an $S^2$ since, for all values of the $D=4$ scalars, the relevant metric functions behave as
\begin{equation}
e^{2\phi - \varphi} \Delta_2^{-1} \, \sin^2\alpha \,  \xrightarrow[\alpha \rightarrow 0]{} \,  \alpha^2 +{\cal O}(\alpha^4) \; , \quad 
e^{2\phi - \varphi} \Delta_2^{-1} \, \sin^2\alpha \,  \xrightarrow[\alpha \rightarrow \pi]{} \,  (\pi-\alpha)^2 +{\cal O}( (\pi-\alpha)^4) ,
\end{equation}
at both endpoints of the chosen range (\ref{eq:ranges}) for $\alpha$. Moreover, this $S^2$ smooth if the period of $\psi$ is chosen as in (\ref{eq:ranges}). Similarly, the bundle over $\mathbb{CP}^2$ is also smooth because
\begin{equation}
e^{2\phi - \varphi} X \Delta_1^{-1} \, \sin^2\alpha \,  \xrightarrow[\alpha \rightarrow 0]{} \,  \alpha^2 +{\cal O}(\alpha^4) \; ,  \quad 
e^{2\phi - \varphi} X \Delta_1^{-1} \, \sin^2\alpha \,  \xrightarrow[\alpha \rightarrow \pi]{} \,  (\pi-\alpha)^2 +{\cal O}( (\pi-\alpha)^4) .
\end{equation}
The easiest way to see that the total space of the bundle is indeed $S^6$ is by noting that the local family of internal metrics in (\ref{KKSU3sectorinIIA}) can be continuously deformed to the usual sine-cone metric (\ref{eq:sineconeS6}) by restricting the $D=4$ scalars to the line
\begin{equation} \label{eq:SO7subtrunc}
\varphi=\phi \; , \qquad 
\chi=a=\zeta=\tilde{\zeta}=0 
\end{equation}
within the scalar manifold (\ref{ScalManN=2}). The warp factor, dilaton and form potentials also remain smooth for all values of the $D=4$ scalars.

The metric in (\ref{KKSU3sectorinIIA}) displays an $\textrm{SU}(3) \times \textrm{U}(1)$ isometry group, where the first factor corresponds to the isometry of the Fubini-Study metric on $\mathbb{CP}^2$, and the U(1) is generated by the vector $\partial_\psi$ dual to $\bm{\eta}$. For $\zeta  \neq 0 $ or $ \tilde \zeta \neq 0$, this U(1) is broken by the dependence of the supergravity forms on the Sasaki-Einstein two-form $\bm{\Omega}$, which carries U(1) charge. The generic symmetry of the complete  IIA configuration (\ref{KKSU3sectorinIIA}) is thus SU(3), in agreement with the symmetry of the sector of $D=4$ ISO(7) supergravity from which it uplifts. When the $D=4$ scalars are restricted to certain loci of (\ref{ScalManN=2}), symmetry enhancements occur. As we have already noted, configurations (\ref{KKSU3sectorinIIA}) with
\begin{eqnarray} \label{eq:SU3U1configs}
\zeta = \tilde \zeta = 0 
\end{eqnarray}
preserve $\textrm{SU}(3) \times \textrm{U}(1)$. The latter acts with cohomogeneity-one. When
\begin{eqnarray} \label{eq:SO6configs}
\chi =  \zeta = \tilde \zeta = 0 \; ,
\end{eqnarray}
the symmetry is enhanced to SO$(6)_+$, which again acts with cohomogeneity-one. For
\begin{eqnarray} \label{eq:SU3toG2}
\phi = \varphi \; , \qquad 
\tilde \zeta = 2\chi \; , \qquad 
a= \zeta = 0 \; ,
\end{eqnarray}
the symmetry is enhanced from SU(3) to G$_2$, which now acts homogeneously (see subsection \ref{subsec:G2fromSU3}). Finally, the symmetry is enhanced to a homogeneous SO$(7)_+$ when the $D=4$ scalars take values on the locus (\ref{eq:SO7subtrunc}). The loci (\ref{eq:SO7subtrunc})--(\ref{eq:SU3toG2}) also define the sectors of ISO(7) supergravity with the same symmetry, see section 3.5 of \cite{Guarino:2015qaa}. Therefore, the evaluation of equation (\ref{KKSU3sectorinIIA}) on these loci together with appropriate restrictions of the $D=4$ $p$-forms determines the truncation of massive IIA to specific $D=4$ sectors. We will discuss some of these truncations in more detail in section \ref{sec:urthertruncs}.

Since supersymmetry is preserved in the uplifting process, the type IIA configuration (\ref{KKSU3sectorinIIA})  is $8/32$--supersymmetric for generic values of the $D=4$ fields, reflecting the  $\cN=2$ supersymmetry of the SU(3)-invariant sector of ISO(7) supergravity. Supersymmetry in (\ref{KKSU3sectorinIIA}) is realised by the presence of a global $\textrm{SU}(3)  \times \textrm{SU}(3)$--structure, or local SU(2)--structure, on the internal $S^6$ of the type discussed in \cite{Lust:2009zb} (see also \cite{Petrini:2009ur,Lust:2009mb}). Since  (\ref{KKSU3sectorinIIA}) only depends on the Sasaki-Einstein structure of $S^5$, these truncation formulae are still locally valid if $S^5$ is replaced with any Sasaki-Einstein manifold. In other words, massive type IIA supergravity can be consistently truncated on a local geometry of the form (\ref{KKSU3sectorinIIA}), based on an arbitrary Sasaki-Einstein five-manifold, to the $D=4$ $\cN=2$ supergravity theory of section 3 of \cite{Guarino:2015qaa}. In this more general setting, the internal geometry will still be that of an $S^2$ bundle over the local K\"ahler-Einstein base of the Sasaki-Einstein manifold. The total space of the bundle will no longer correspond to the six-sphere, and will typicaly display orbifold-type singularities.

\section{Further truncations} \label{sec:urthertruncs}

It is instructive to particularise the above SU(3)--invariant consistent truncation formulae to further subsectors of the $D=4$ theory with smaller field content and larger symmetry.

%%%%%%%%%%%%%%%
%%%%%%%%%%%%%%%
\subsection{Truncation to the G$_2$ sector} 
\label{subsec:G2fromSU3}

%%%%%%%%%%%%%%%
%%%%%%%%%%%%%%%

The G$_2$-invariant sector of $D=4$ ISO(7) supergravity was discussed in \cite{Guarino:2015qaa}, and the explicit truncation from type IIA, in \cite{Guarino:2015vca}. Here, we will recover the latter truncation from the SU(3)--invariant truncation. 

The $D=4$ G$_2$ sector is recovered from the SU(3) sector by setting to zero all vectors and two-form potentials in (\ref{fieldContentHierarchy}), identifying the scalars as in (\ref{eq:SU3toG2}), and the three-form potentials as $C^0 = C^1 \equiv C$.
Bringing these identifications to (\ref{KKSU3sectorinIIA}), we find that all the $\alpha$ dependence combines with the forms $\bm{J}, \bm{\Omega}, \bm{\eta}$ into the combinations (\ref{NKintermsof SE5Forms}) that determine the homogeneous nearly-K\"ahler structure ${\cal J}$, $\Upomega$ on $S^6$. Specifically, (\ref{KKSU3sectorinIIA}) reduces to
{\setlength\arraycolsep{0pt}
\begin{eqnarray} \label{G2embeddingGeom}
&& d \hat{s}_{10}^2 = e^{\frac{3}{4} \varphi} \big( 1+e^{2 \varphi} \chi^2 \big)^{\frac{3}{4}}  ds^2_4 + g^{-2} e^{-\frac{1}{4} \varphi} \big( 1+e^{2 \varphi} \chi^2 \big)^{-\frac{1}{4}}   ds^2(S^6) \; , \nonumber \\[5pt]
&& e^{\hat \phi} = e^{\frac{5}{2} \varphi} \big( 1+e^{2 \varphi} \chi^2 \big)^{-\frac32}  \; , \nonumber \\[5pt]
&&  \hat A_\3 = C + g^{-3} \chi \, \textrm{Im} \, \Upomega  \; , \qquad
 \hat B_\2 = g^{-2} \, e^{2 \varphi} \chi  \big( 1+e^{2 \varphi} \chi^2 \big)^{-1}   \,  {\cal J}  \; , \qquad   
 \hat A_\1 = 0 \; .
\end{eqnarray}
}These agree with the formulae for the consistent truncation to the G$_2$--invariant sector given in (4.3) of  \cite{Guarino:2015vca}. Likewise, the constant-scalar field strengths (\ref{KKSU3FieldStrengths}) reduce to the field strengths given in (4.4) of  \cite{Guarino:2015vca}, after dropping the terms in the latter containing derivatives of scalars.

%%%%%%%%%%%%%%%%%
%%%%%%%%%%%%%%%%%

\subsection{Dilatons}
\label{sec:dilatons}
%%%%%%%%%%%%%%%%%
%%%%%%%%%%%%%%%%%

Here we give an illustrative example of how the dualisation conditions of  \cite{Guarino:2015qaa} can be used to simplify not only the Freund-Rubin term, but also the three-form field strength contributions. For simplicity, we focus on the consistent subsector of (\ref{fieldContentHierarchy}) containing only the dilatons
$\varphi$, $\phi$, the two form $B_1$ and the three-forms $C^0$, $C^1$, with all other fields set to zero. This is a consistent subsector both of the SO$(6)_+$ invariant sector and of the neutral scalar model of \cite{Guarino:2015jca}. From (3.14), (3.15) of  \cite{Guarino:2015qaa}, the field strengths in this subsector are
\begin{eqnarray} \label{eq:subsectorFS}
H_{\3 1} = dB_1 + 2 g (C^1 - C^0) \; ,  \qquad H_{\4}^0 = dC^0\; ,  \qquad H_{\4}^1 = dC^1 \; .
\end{eqnarray}
These are subject to the dualisation conditions
\begin{equation} \label{eq:subsectorFSDual}
{H_{\3 1}} =   2 * \big(  d \varphi - d\phi   \big) \; , \quad 
H_{\4}^{0} =  g \,  \big( 6 \, e^{2 \phi - \varphi } - e^{4 \phi - 3\varphi }   \big)   \, \textrm{vol}_{4} \; , \quad 
H_{\4}^{1} =  g \big( 4 \, e^{\varphi } +  e^{2 \phi - \varphi }   \big)  \, \textrm{vol}_{4} \; ,
\end{equation}
which arise as a particular case of (3.18), (3.19) of \cite{Guarino:2015qaa}.

In this sector, the functions (\ref{Delta1}), (\ref{Delta2}) simplify to 
\begin{eqnarray}
\Delta_1 = \Delta_2 \equiv \Delta =  e^{\varphi}  \sin^2 \alpha  + e^{2\phi-\varphi}  \cos^2 \alpha \; , 
\end{eqnarray}
and the type IIA embedding (\ref{KKSU3sectorinIIA}) to
{\setlength\arraycolsep{0pt}
\begin{eqnarray} \label{KKDilatons}
&& d\hat{s}_{10}^2 =  e^{\frac18 (2\phi-\varphi)}  \Delta^{5/8}  \Big[ \; ds_4^2   + g^{-2} e^{-2\phi+\varphi}   d\alpha^2  +  g^{-2} \Delta^{-1}   \sin^2 \alpha   \, d\tilde{s}^2 ( S^5 )  \Big] \; ,  \nonumber \\[5pt]
&& e^{\hat \phi}  =  e^{\frac14 (6\phi +5\varphi)}  \Delta^{-1/4} \; , \nonumber \\[5pt]
&& \hat  A_\3 = \cos^2 \alpha \, C^0  + \sin^2 \alpha \, C^1    -g^{-1} \sin \alpha \cos \alpha \, B_1  \wedge d \alpha  
 \; , \quad
\hat  B_\2 = 0 \; , \quad
\hat  A_\1 = 0 \; .
\end{eqnarray}
}Here, $d\tilde{s}^2 ( S^5 )$ is the round, SO(6)-invariant metric on the foliating $S^5$, normalised so that the Ricci tensor equals 4 times the metric. Now, with the help of the definitions (\ref{eq:subsectorFS}), the four-form field strength corresponding to the configuration (\ref{KKDilatons}) can be written as
\begin{eqnarray}
\hat F_\4 =   H_{\4}^0 \, \cos^2\alpha + H_{\4}^1 \, \sin^2\alpha -g^{-1} \sin\alpha \cos \alpha  \, H_{\3 1} \wedge d\alpha \; .
\end{eqnarray}
Finally, the dualisation conditions (\ref{eq:subsectorFSDual}) can be used to rewrite $\hat F_\4$ in terms of independent $D=4$ fields only ($\phi$, $\varphi$ and the metric, through the Hodge dual), as
\begin{eqnarray}
\hat F_\4 &= & g  \Big[   \big( 6 \, e^{2 \phi - \varphi } - e^{4 \phi - 3\varphi }   \big)  \, \cos^2\alpha +  \big( 4 \, e^{\varphi } +  e^{2 \phi - \varphi }   \big)  \, \sin^2\alpha  \Big] \textrm{vol}_4 \nonumber \\
&& +2g^{-1} \sin\alpha \cos \alpha  \, d\alpha \wedge * \big(  d \varphi - d\phi   \big)  \; .
\end{eqnarray}
The Freund-Rubin term here is a particularisation of (\ref{USU3}). Terms like  $ * \big(  d \varphi - d\phi   \big)$ in $\hat F_\4$ have long been known to enter similar spherical truncation formulae, see {\it e.g.} (3.1) of \cite{Cvetic:1999xx} or the formulae in \cite{Cvetic:2000dm}. Such term has been obtained here using the systematics of the duality hierarchy.

It is futher consistent to set $\varphi = \phi$ in the above formulae. The resulting expressions correspond to the truncation of IIA to the SO$(7)_+$--invariant sector of ISO(7) supergravity.

%%%%%%%%%%%%%%%%%
%%%%%%%%%%%%%%%%%
\section{New AdS$_4$ solutions of massive type IIA} 
\label{sec:AdSsolutions}
%%%%%%%%%%%%%%%%%
%%%%%%%%%%%%%%%%%

We will now use the explicit formulae (\ref{KKSU3sectorinIIA}), (\ref{KKSU3FieldStrengths}) for the ten-dimensional uplift of the SU(3)--invariant sector of $D=4$ (dyonic) ISO(7) supergravity, to obtain sourceless solutions of massive type IIA supergravity. All $D=4$ critical points with at least SU(3) symmetry are AdS, and their location in the scalar manifold (\ref{ScalManN=2}) can be found in table 3 of \cite{Guarino:2015qaa}. Evaluating (\ref{KKSU3sectorinIIA}), (\ref{KKSU3FieldStrengths}) at the scalar locations given in that table, we find classical solutions of massive type IIA of the form $\textrm{AdS}_4 \times S^6$, with the product possibly warped and with various metrics on $S^6$. The supersymmetry (or lack thereof) and bosonic symmetry of the $D=4$ critical points is carried over into the ten-dimensional solutions. The former, if present, endows the $S^6$ with a particular $G$-structure. The latter manifests itself as the subgroup of the isometry group of the relevant metric on $S^6$ that also preserves the IIA form fields. The uplift of critical points with symmetry group $G_0$ such that G$_2 \subset G_0$ leads to homogeneous solutions, in agreement with the discussion in section \ref{sec:Regularity}. On the other hand, critical points with $\textrm{SU}(3) \subset G_0 \subset \textrm{G}_2$, lead to cohomogeneity-one solutions in ten dimensions. By the analysis of section \ref{sec:Regularity}, all these solutions are smooth. 

The $\cN=1$ SU(3) solution (\ref{SU3IIASolutionRescaled}) and the $\cN=0$ SO$(6)_+$ solution (\ref{SO6SolIIARescaled}) are new. The uplift of the $D=4$ $\cN=2$ point with $\textrm{SU}(3) \times \textrm{U}(1)$ symmetry is omitted here, as it was already given in \cite{Guarino:2015jca}. All other solutions have already appeared in the literature, as they have been previously obtained by other methods. In any case, we have checked that all these configurations solve the massive type IIA field equations, see appendix \ref{SU3InvAdS4Sols}. This provides a double-check on our uplifting formulae and on the consistency of the truncation. 

The solutions below are presented in ten-dimensional Einstein frame. We have rescaled the external four-dimensional metric so that AdS$_4$ is always unit radius. The metrics thus have the local form
\begin{eqnarray} \label{GenericFormMetric}
d\hat{s}^2_{10} = e^{2A} \Big( ds^2( \textrm{AdS}_4) + d s_6^2 \Big) \; .
\end{eqnarray}
From (\ref{KKSU3sectorinIIA}) and (\ref{UintermsofVCritical}), the warp factor and Freund-Rubin term become
\begin{equation} \label{eq:WarpFactorandFR}
e^{2A} \equiv -6 \, e^{\frac18 (2\phi-\varphi)} X^{1/4}  \Delta_1^{1/2}  \Delta_2^{1/8} \, V_0^{-1} \quad , \qquad 
U_\textrm{rescaled} = -12 \,  g^{-1} \,  V_0^{-1} \; , 
\end{equation}
where $V_0 < 0 $ is the value of the SU(3)-invariant scalar potential, (3.11) of \cite{Guarino:2015qaa}, at each critical point. All the solutions depend on the $D=4$ couplings $g$ and $m$ through combinations $L$ and $e^{\phi_0}$ that we define for each solution on an individual basis. Although the uplifting formulae  (\ref{KKSU3sectorinIIA}) for the metric, dilaton and form potentials only depend on $g$, and $m$ only enters the embedding through the field strengths (\ref{Romans=m}), (\ref{KKSU3FieldStrengths}), all the fields in the solutions below develop a dependence on both $g$ and $m$ (via $L$ and $e^{\phi_0}$). The reason for this is that the position in the salar space (\ref{ScalManN=2}) of the $D=4$ critical points depends on both couplings (as $\sim g^{\frac13} m^{-\frac13}$), see table 3 of \cite{Guarino:2015qaa}.

We comment on generalisations of these solutions and, for the non-supersymmetric ones, on their stability against fluctuations contained in the $D=4$ $\cN=8$ dyonic ISO(7) supergravity. We conclude with a flux quantisation analysis. The latter shows that all stable solutions are also good massive type IIA string theory backgrounds. See table \ref{Table:SU3Points} in the introduction for a summary.

\subsection{Supersymmetric solutions} \label{subsec:susysols}

Recall that dyonic ISO(7) supergravity contains supersymmetric points with ($\cN=2$, $\textrm{SU}(3) \times \textrm{U}(1)$),  ($\cN=1$, $\textrm{SU}(3)$) and ($\cN=1$, G$_2$) residual supersymmetry and bosonic symmetry. The massive IIA solution corresponding to the $\cN=2$ point has already been given in \cite{Guarino:2015jca}. For that reason, we omit the full solution here and merely give a set of potentials for the internal field strengths\footnote{The field strengths of the solution in \cite{Guarino:2015jca} were in fact computed from this set of potentials, and then double-ckeched with (\ref{KKSU3FieldStrengths}). A set of potentials has also been recently given in \cite{Fluder:2015eoa}.}:
\begin{eqnarray} \label{SU3U1lIIAPotentials}
&& L^{-3} e^{\frac14 \phi_0}  \hat A_\3 = 6\sqrt{3}  \, \frac{  \sin^4 \alpha }{ 3 + \cos 2\alpha }  \ \bm{J}   \wedge \bm{\eta} \; , 
\nonumber \\[5pt]
&& L^{-2} e^{-\frac12 \phi_0}  \hat B_\2 =   -6  \sqrt{2}  \ \frac{  \sin^2 \alpha \cos \alpha  }{3 + \cos 2\alpha  }  \ \bm{J}     - \frac{3}{\sqrt{2}} \,  \sin \alpha \ d\alpha \wedge \bm{\eta}   \; , 
\nonumber \\[5pt]
&& L^{-1} e^{\frac34 \phi_0}  \hat A_\1 =   \sqrt{6}  \ \frac{  \sin^2 \alpha \cos \alpha  }{ 5 + \cos 2\alpha}  \,  \bm{\eta}   \; . 
\end{eqnarray}
These follow from particularising (\ref{KKSU3sectorinIIA}) to the $\cN=2$, $\textrm{SU}(3) \times \textrm{U}(1)$ critical point, and then performing a gauge transformation to remove the constant $D=4$ St\"uckelberg scalar $a$. The contributions from the external three-forms $C^0$ and $C^1$ have been excluded here. Following \cite{Guarino:2015jca}, we have defined the constants $L^2 \equiv 2^{-\frac{5}{8}} \, 3^{-1} \, g^{-\frac{25}{12}} \,  m^{\frac{1}{12}}$ and $e^{\phi_0} \equiv  2^{\frac{1}{4}} \, g^{\frac{5}{6}} \,  m^{-\frac{5}{6}}$, so that, for this solution, $g = 2^{-\frac{3}{10}} \, 3^{-\frac12} \, L^{-1} \, e^{-\frac{1}{20} \phi_0}$ and $\hat F_\0 \equiv m = 3^{-\frac12} \, L^{-1} \, e^{-\frac{5}{4} \phi_0}$.

The $\cN=1$ SU(3) extremum gives rise to a new $\cN=1$ solution of massive type IIA supergravity. It reads
{\setlength\arraycolsep{2pt}
\begin{eqnarray}  \label{SU3IIASolutionRescaled}
d \hat{s}_{10}^2 & = & L^2 \, \big( 3 + \cos 2\alpha \big)^{1/2} \big( 2 + \cos 2\alpha \big)^{1/8}   \Big[  \, ds^2(\textrm{AdS}_4)   \nonumber \\
&& \quad \qquad  + \frac65 \,  d\alpha^2 +  \frac{24 \sin^2 \alpha}{ 5( 3 + \cos 2\alpha )} \,ds^2 ( \mathbb{CP}^2 )  +   \frac{ 18 \sin^2 \alpha}{5 \big( 2 + \cos 2\alpha \big)} \,  \bm{\eta}^2  \Big] \; , \nonumber \\[12pt]
 e^{\hat \phi} &=& e^{\phi_0} \frac{  \big( 2 + \cos 2\alpha \big)^{3/4}  }{ 3 + \cos 2\alpha } \;  , \nonumber  \\[12pt]
L^{-3} e^{\frac14 \phi_0}   \hat F_\4 &=&  \frac{3\sqrt{5}}{\sqrt{2}} \, \textrm{vol} ( \textrm{AdS}_4 ) - \frac{96\sqrt{6}}{25}  \,  \frac{  4 +  3\cos 2\alpha }{   \big( 3 + \cos 2\alpha \big)^2   } \, \sin^4 \alpha \ \textrm{vol} (  \mathbb{CP}^2) \nonumber \\
&&  -\frac{72\sqrt{6}}{25} \, \frac{  6 +  \cos 2\alpha }{   \big( 2 + \cos 2\alpha \big)  \big( 3 + \cos 2\alpha \big)   } \sin^3 \alpha \cos \alpha \ \bm{J} \wedge d\alpha \wedge \bm{\eta}  \;   \nonumber \\
&&  +\frac{648\sqrt{2}}{25}  \frac{ \sin^4 \alpha }{   \big( 2 + \cos 2\alpha \big)  \big( 3 + \cos 2\alpha \big)   }  \, \textrm{Im} \, \bm{\Omega}  \wedge d\alpha \wedge \bm{\eta}  \; , 
\nonumber \\[12pt]
L^{-2} e^{-\frac12 \phi_0}  \hat H_\3 & =&   -  \frac{ 96 }{ 5\sqrt{5} }   \,  \frac{  \sin^3 \alpha }{   \big( 3 + \cos 2\alpha \big)^2   }  \ \bm{J} \wedge d\alpha  \;  \nonumber \\
&& - \frac{24 \sqrt{3}}{5\sqrt{5} } \frac{ 11+ \cos 2\alpha  }{ \big( 3 + \cos 2\alpha \big)^2 } \,  \sin^2\alpha \cos \alpha  \,\, \textrm{Im} \, \bm{\Omega}  \wedge d\alpha  - \frac{72 \sqrt{3}}{5\sqrt{5} }  \frac{ \sin^3 \alpha   }{ 3 + \cos 2\alpha  } \, \,  \textrm{Re} \, \bm{\Omega}  \wedge \bm{\eta} \; , 
\nonumber \\[12pt]
L^{-1} e^{\frac34 \phi_0}  \hat F_\2 & =&   - \frac{ 4 \sqrt{6} }{ \sqrt{5} }  \, \frac{  \sin^2 \alpha \cos \alpha  }{   \big( 2 + \cos 2\alpha \big)  \big( 3 + \cos 2\alpha \big)   }  \ \bm{J}  
\,  - \frac{3 \sqrt{6} }{\sqrt{5} } \,  \frac{ \sin \alpha \cos 2\alpha   }{   \big( 2 + \cos 2\alpha \big)^2   }   \, \ d\alpha \wedge \bm{\eta}  \nonumber \\
&& -  \frac{12 \sqrt{2} }{\sqrt{5} } \frac{ \sin^3 \alpha   }{ 3 + \cos 2\alpha  } \, \, \textrm{Im} \, \bm{\Omega}  \; .
\end{eqnarray}
}The solution depends on the $D=4$ gauge coupling constants $g$, $m$ through the combinations $L$, $e^{\phi_0}$, which are now defined as $L^2 \equiv 2^{-\frac{7}{2}} \, 3^{-1} \, 5^{\frac{11}{8}} \, g^{-\frac{25}{12}} \,  m^{\frac{1}{12}}$ and $e^{\phi_0} \equiv  2 \cdot  5^{\frac{1}{4}} \, g^{\frac{5}{6}} \,  m^{-\frac{5}{6}}$, so that $g = 2^{-\frac{17}{10}} \, 3^{-\frac12} \, 5^{\frac{7}{10}} \, L^{-1} \, e^{-\frac{1}{20} \phi_0}$ and $\hat F_\0 \equiv m = 2^{-\frac{1}{2}} \, 3^{-\frac12} \, 5 \, L^{-1} \, e^{-\frac{5}{4} \phi_0}$. A set of internal potentials for this solution follows from (\ref{KKSU3sectorinIIA}):
{\setlength\arraycolsep{2pt}
\begin{eqnarray}  \label{SU3IIASolutionPotentials}
L^{-3} e^{\frac14 \phi_0}   \hat A_\3  &=&   -  \frac{ 48 \sqrt{6} }{ 25 }   \,  \frac{  \sin^4 \alpha }{ 3 + \cos 2\alpha    }  \ \bm{J} \wedge \bm{\eta}   + \frac{72 \sqrt{2} }{25}  \,  \sin^2\alpha \, \, \textrm{Re} \, \bm{\Omega}  \wedge d\alpha  \nonumber \\
&&  - \frac{288 \sqrt{2}}{25 } \,   \frac{ \sin^3 \alpha   \cos \alpha }{ 3 + \cos 2\alpha  } \,   \, \textrm{Im} \, \bm{\Omega}  \wedge \bm{\eta} \; , 
\nonumber \\[10pt]
L^{-2} e^{-\frac12 \phi_0}  \hat B_\2 &=&  \frac{ 24 }{5 \sqrt{5} }  \, \frac{  \sin^2 \alpha \cos \alpha  }{ 3 + \cos 2\alpha }  \ \bm{J}  
 + \frac{6 }{5\sqrt{5} } \, \sin \alpha \, \ d\alpha \wedge \bm{\eta}  -  \frac{24 \sqrt{3} }{5\sqrt{5} } \frac{ \sin^3 \alpha   }{ 3 + \cos 2\alpha  } \, \, \textrm{Im} \, \bm{\Omega} \; ,
\nonumber \\[10pt]
L^{-1} e^{\frac34 \phi_0}  \hat A_\1  &=&   - \frac{2 \sqrt{6} }{\sqrt{5} } \,  \frac{ \sin^2 \alpha \cos \alpha   }{  2 + \cos 2\alpha  }   \,  \bm{\eta}  \; .
\end{eqnarray}
}In fact, we first computed the metric and dilaton in (\ref{SU3IIASolutionRescaled}) and the form potentials in (\ref{SU3IIASolutionPotentials}) using the formulae (\ref{KKSU3sectorinIIA}). The field strengths in (\ref{SU3IIASolutionRescaled}) were then computed from (\ref{SU3IIASolutionPotentials}) using their type IIA definitions (see (A.4) of \cite{Guarino:2015vca} for our conventions), and then double-ckeched with (\ref{KKSU3FieldStrengths}). In (\ref{SU3IIASolutionPotentials}), as in (\ref{SU3U1lIIAPotentials}), the constant $D=4$ scalar $a$ has been gauged away.

Recall from section \ref{subsec:SU3UpliftSubsec} that the Fubini-Study metric $ds^2 ( \mathbb{CP}^2 )$ that appears in  (\ref{SU3IIASolutionRescaled}) is normalised so that the Ricci tensor equals 6 times the metric, and $\bm{J}$, $\bm{\Omega}$, $\bm{\eta}$ are the forms that characterise the Sasaki-Einstein structure of a foliating $S^5$ inside $S^6$. In particular, $\bm{J}$ can be regarded as the K\"ahler form of $\mathbb{CP}^2 $. The Hopf fiber of the $S^5$ is  stretched\footnote{The fiber is stretched in the sense that, at the $\cN=1$, SU(3) critical point,  $X^{-1} \Delta_1 \Delta_2^{-1} \geq 1$ for all $\alpha$. In contrast, the fiber of the solution (12) of \cite{Guarino:2015jca} is squashed since, at the $\cN=2$, $\textrm{SU}(3) \times \textrm{U}(1)$ critical point, $X^{-1} \Delta_1 \Delta_2^{-1}\leq 1$ for all $\alpha$.}, and the vector $\partial_\psi$, dual to the one-form $\bm{\eta} = d\psi + \sigma$ along the fiber, is  a U(1) Killing vector of the metric. The isometry is therefore $\textrm{SU}(3) \times \textrm{U}(1)$, and acts with cohomogeneity-one. However, the dependence of the supergravity forms on $\bm{\Omega}$ breaks the symmetry of the complete IIA solution to SU(3), in agreement with the symmetry of the $D=4$ critical point. The $\cN=1$ supersymmetry of the solution (\ref{SU3IIASolutionRescaled}) is reflected by the presence of a global $\textrm{SU}(3)  \times \textrm{SU}(3)$--structure, or local SU(2)--structure, of the type discussed in \cite{Lust:2009zb} (see also \cite{Lust:2009mb}). Since the solution only depends on the Sasaki-Einstein structure of $S^5$, a more general class of $\cN=1$ solutions of massive type IIA can be obtained from (\ref{SU3IIASolutionRescaled}) by replacing $S^5$ with an arbitray Sasaki-Einstein five-manifold. The resulting solutions will have other topologies than $S^6$ and will typically become singular.

The $\cN=1$ G$_2$ critical point uplifts via (\ref{KKSU3sectorinIIA}), (\ref{KKSU3FieldStrengths}), or equivalently  (\ref{G2embeddingGeom}), to the solution
\begin{eqnarray} \label{G2N=1SolIIARescaled}
&& d \hat{s}_{10}^2 = L^2 \, \Big(  ds^2(\textrm{AdS}_4) + \tfrac{12}{5} \,  ds^2(S^6) \Big)\; , \qquad 
e^{\hat \phi} = e^{\phi_0}  \; , \qquad 
L^{-1} e^{\frac34 \phi_0}   \hat F_{(2)} = -\tfrac{ \sqrt{3} }{ \sqrt{5}}  \, {\cal J}   \; ,  \nonumber\\
&& L^{-3} e^{\frac14 \phi_0}  \hat F_{(4)} =  \tfrac{3\sqrt{15}}{4}     \,  \textrm{vol}(\textrm{AdS}_4) +\tfrac{54}{25}   \, {\cal J} \wedge {\cal J}  \; ,  \qquad 
L^{-2} e^{-\frac12 \phi_0}   \hat H_{(3)} =- \tfrac{12\sqrt{3}}{5\sqrt{5}} \, \textrm{Re} \, \Upomega  \; , 
\end{eqnarray}
first found by Behrndt and Cvetic in \cite{Behrndt:2004km} using $G$-structure techniques. We have defined the constants $L^2 \equiv 2^{-\frac{43}{12}} \, 3^{-\frac{5}{8}}\, 5^{\frac{11}{8}} \, g^{-\frac{25}{12}} \,  m^{\frac{1}{12}}$ and $e^{\phi_0} \equiv  2^{-\frac{1}{6}} \,15^{\frac{1}{4}} \, g^{\frac{5}{6}} \,  m^{-\frac{5}{6}}$, so that $g = 2^{-\frac{9}{5}} \, 3^{-\frac{3}{10}} \, 5^{\frac{7}{10}} \, L^{-1} \, e^{-\frac{1}{20} \phi_0}$ and $\hat F_\0 \equiv m = \frac{5}{4} \, L^{-1} \, e^{-\frac{5}{4} \phi_0}$. The metric $ds^2(S^6)$ is the Einstein metric on the unit radius $S^6$ (normalised so that the Ricci tensor equals 5 times the metric), now regarded as the homogeneous space G$_2/$SU(3). The real two-form ${\cal J}$ and holomorphic $(3,0)$-form $\Upomega$ are G$_2$--invariant, and endow $S^6$ with its natural homogeneous nearly-K\"ahler structure, see appendix \ref{eq:SplitS6intoS5}. As noted in \cite{Behrndt:2004km}, (\ref{G2N=1SolIIARescaled}) can be promoted to a class of $\cN=1$ solutions by replacing $S^6$ with an arbitrary nearly-K\"ahler manifold.

\subsection{Non-supersymmetric solutions} \label{subsec:nonsusysols}

We now turn to discuss the uplift of the non-supersymmetric critical points of dyonic ISO(7) supergravity with at least SU(3) symmetry. In this sector, $\cN=0$ points with residual SO$(7)_+$, G$_2$ and SO$(6)_+$ symmetry are known analytically. In addition, two points with SU(3) symmetry are known numerically. We will focus on the analytic points --only a few comments about the numerical ones will be made at the end of this section.

The SO$(6)_+$ critical point gives rise to a new non-supersymmetric solution of massive type IIA:
\begin{eqnarray} \label{SO6SolIIARescaled}
&& d \hat{s}_{10}^2 = L^2  \big(5 + 3 \cos 2\alpha \big)^{\frac{5}{8}}  \Big[  ds^2(\textrm{AdS}_4) +  d\alpha^2 + \frac{8\sin^2 \alpha}{ 5+3\cos 2\alpha}   d\tilde{s}^2(S^5) \Big] \; ,\\
&& e^{\hat \phi} = e^{\phi_0} \big(5 + 3 \cos 2\alpha \big)^{-\frac{1}{4}}  \; , \quad 
 \hat F_{(4)} =  8L^3 e^{-\frac14 \phi_0} \,  \textrm{vol}(\textrm{AdS}_4) \; ,  \quad 
\hat H_{(3)} = 0  \; , \quad
\hat F_{(2)} = 0   \; .  \nonumber 
\end{eqnarray}
Here, $d\tilde{s}^2(S^5)$ is the round, Einstein metric on the unit radius $S^5 = \textrm{SO}(6)/\textrm{SO}(5)$, normalised so that the Ricci tensor equals 4 times the metric. We have defined the constants  $L^2 \equiv 2^{-\frac{7}{3}} \, g^{-\frac{25}{12}} \,  m^{\frac{1}{12}}$ and $e^{\phi_0} \equiv  2^{\frac{4}{3}} \, g^{\frac{5}{6}} \,  m^{-\frac{5}{6}}$, so that $g = 2^{-\frac{11}{10}} \, L^{-1} \, e^{-\frac{1}{20} \phi_0}$ and $\hat F_\0 \equiv m = \sqrt{2} \, L^{-1} \, e^{-\frac{5}{4} \phi_0}$. The group SO(6) acts by isometries with co\-ho\-mo\-ge\-neity-one. 

This solution is unstable, given that the spectrum \cite{DallAgata:2011aa} of the SO$(6)_+$ point of dyonic ISO(7) supergravity contains BF-bound-violating modes. This spectrum coincides, within their $\cN=8$ truncations, with that \cite{Bobev:2010ib} of the SU$(4)_-$ point \cite{Warner:1983vz} of $D=4$ SO(8)-gauged supergravity \cite{deWit:1982ig}. The instability of the solution (\ref{SO6SolIIARescaled}) is thus formally analogous to that \cite{Bobev:2010ib} of the Pope-Warner solution \cite{Pope:1984bd} of $D=11$ supergravity that arises from uplift on $S^7$ of the SU$(4)_-$ point. The actual details on how the instability affects the massive IIA SO$(6)_+$ solution (\ref{SO6SolIIARescaled}) and the $D=11$ Pope-Warner SU$(4)_-$ solution will differ, given the very different structure of the solutions. We also note that a wider class of non-supersymmetric solutions with topologies different from $S^6$, and possibly singular, may be obtained from (\ref{SO6SolIIARescaled}) simply by replacing the round $S^5$ with any Einstein five-manifold $M_5$. The resulting symmetry will then be that of $M_5$, and the stability issue would need to be readdressed on a case-by-case basis. Little hope should be harboured about the solution (\ref{SO6SolIIARescaled}) becoming stable for other Einstein manifolds $M_5$: the instabilities were found to persist \cite{Pilch:2013gda} for similar generalisations of the Pope-Warner solution.

The solutions corresponding to the uplift of the $\cN=0$  SO(7)$_+$ and G$_2$ points were previously known, although the stability comments we present here are new. The SO(7)$_+$ critical point gives rise to a homogeneous, Freund-Rubin type solution first found by Romans \cite{Romans:1985tz}. In our conventions, this solution reads
\begin{eqnarray} \label{SO7SolIIA}
&& d \hat{s}_{10}^2 =L^2 \, \Big(  ds^2(\textrm{AdS}_4) + \tfrac52  \, ds^2(S^6) \Big)\; , \quad 
e^{\hat \phi} = e^{\phi_0}  \; , \nonumber \\
&&  \hat F_{(4)} =  \sqrt{10} \, L^3 e^{-\frac14 \phi_0}  \,  \textrm{vol}(\textrm{AdS}_4) \; ,  \quad \hat H_{(3)} = 0  \; , \quad
\hat F_{(2)} = 0   \; .
\end{eqnarray}
Here, $ds^2(S^6)$ is the round, Einstein metric on the unit radius $S^6 = \textrm{SO}(7)/\textrm{SO}(6)$, with normalisation such that the Ricci tensor equals 5 times the metric. We have defined the constants $L^2 \equiv 2 \cdot 5^{-\frac{25}{24}} \, g^{-\frac{25}{12}} \,  m^{\frac{1}{12}}$ and $e^{\phi_0} \equiv  5^{\frac{5}{12}} \, g^{\frac{5}{6}} \,  m^{-\frac{5}{6}}$, so that  $g = \sqrt{\frac{2}{5}} \, L^{-1} \, e^{-\frac{1}{20} \phi_0}$ and $\hat F_\0 \equiv m = \sqrt{2} \, L^{-1} \, e^{-\frac{5}{4} \phi_0}$.

The spectrum \cite{DallAgata:2011aa} of the SO$(7)_+$ point of dyonic ISO(7) supergravity contains modes below the BF bound. The solution (\ref{SO7SolIIA}) is, thus, unstable. The SO$(7)_+$ point has the same spectrum, whithin their respective $\cN=8$ theories, as the SO$(7)_\pm$ points of SO(8) supergravity \cite{deWit:1982ig}. In this sense, the instability of the SO$(7)_+$ massive IIA solution (\ref{SO7SolIIA}) is qualitatively similar to the instability of the de Wit-Nicolai \cite{deWit:1984va} and Englert \cite{Englert:1982vs} solutions of $D=11$ supergravity, that respectively arise from uplift on $S^7$ of the  SO$(7)_+$ and SO$(7)_-$ critical points of the $D=4$ SO(8) gauging. The precise details about how the instability affects these SO$(7)$--invariant solutions in $D=10$ and $D=11$ may again differ. The unstable modes lie in the $\bm{27}$ of SO(7). Since this representation is irreducible under G$_2$ (in particular, there is no branching into singlets), the unstable modes are truncated out from the G$_2$--invariant, universal nearly-K\"ahler truncation of \cite{KashaniPoor:2007tr,Cassani:2009ck}. This is analogous to the observation \cite{Bobev:2010ib} that the universal M-theory truncation on (skew-whiffed) Sasaki-Einstein manifolds \cite{Gauntlett:2009zw} is also blind to the instability of the Pope-Warner solution  \cite{Pope:1984bd}. See \cite{Gauntlett:2009bh} for the spectrum of this solution within the Sasaki-Einstein truncation of \cite{Gauntlett:2009zw}. In this $D=11$  case, the instability is already visible in larger left-invariant truncations of M-theory on $S^7$  \cite{Cassani:2012pj,Cassani:2011fu}. The latter truncations, however, do not have a massive type IIA on $S^6$ analog. A wider class of non-supersymmetric solutions can be obtained from (\ref{SO7SolIIA}) if $S^6$ is replaced with any six-dimensional Einstein space $M_6$, as noted in \cite{Romans:1985tz}. The symmetry will now be that of $M_6$, and the stability issue should in principle be readdressed on a case-by-case basis.

The $\cN=0$ G$_2$ critical point gives rise to a homogeneous solution first found in \cite{Lust:2008zd} using $G$-structure techniques. In our conventions, it reads
\begin{eqnarray} \label{G2N=0SolIIARescaled}
&& d \hat{s}_{10}^2 = L^2 \, \Big(  ds^2(\textrm{AdS}_4) + 2 \,  ds^2(S^6) \Big)\; , \qquad 
e^{\hat \phi} = e^{\phi_0}  \; , \qquad 
L^{-1} e^{\frac34 \phi_0}   \hat F_{(2)} =  \sqrt{2}  \, {\cal J}   \; ,  \nonumber\\
&& L^{-3} e^{\frac14 \phi_0}  \hat F_{(4)} =  \tfrac{3}{\sqrt{2}}     \,  \textrm{vol}(\textrm{AdS}_4) - \sqrt{6}  \, {\cal J} \wedge {\cal J}  \; ,  \qquad 
L^{-2} e^{-\frac12 \phi_0}   \hat H_{(3)} = 2 \sqrt{3} \, \textrm{Re} \, \Upomega  \; , 
\end{eqnarray}
where $L^2 \equiv 2^{-\frac{11}{6}} \, 3^{\frac{3}{8}} \, g^{-\frac{25}{12}} \,  m^{\frac{1}{12}}$ and $e^{\phi_0} \equiv  2^{\frac{1}{3}} \,3^{\frac{1}{4}} \, g^{\frac{5}{6}} \,  m^{-\frac{5}{6}}$, so that $g = 2^{-\frac{9}{10}} \, 3^{\frac{1}{5}} \, L^{-1} \, e^{-\frac{1}{20} \phi_0}$ and $\hat F_\0 \equiv m = \sqrt{\frac{3}{2}} \, L^{-1} \, e^{-\frac{5}{4} \phi_0}$. The rest of the symbols are like in the supersymmetric solution (\ref{G2N=1SolIIARescaled}). A new observation about this solution is that it is perturbatively stable, at least against $D=4$ perturbations contained either in the $\cN=8$ ISO(7) dyonic theory or in the G$_2$--invariant, nearly-K\"ahler  truncation \cite{KashaniPoor:2007tr,Cassani:2009ck}. See \cite{Borghese:2012qm} for the spectrum of the $\cN=0$, G$_2$-invariant point within (the `massless mode') $D=4$ $\cN=8$ dyonic ISO(7) supergravity and \cite{Guarino:2015qaa} for the spectrum within the `massive mode' nearly-K\"ahler truncation \cite{KashaniPoor:2007tr,Cassani:2009ck}. The solution (\ref{G2N=0SolIIARescaled}) can be generalised by replacing $S^6$ with an arbitrary nearly-K\"ahler manifold. The stability of this more general class of solutions should be readdressed. The $\cN=0$ G$_2$ critical point of dyonic ISO(7) supergravity does not have a counterpart in the purely electric SO(8) gauging \cite{deWit:1982ig}. It does, however, in the dyonic SO(8) gauging \cite{Dall'Agata:2012bb}, but the latter does not uplift to $D=11$ \cite{deWit:2013ija,Lee:2015xga}. For this reason, the G$_2$--invariant massive IIA solution (\ref{G2N=0SolIIARescaled}) does not have a $D=11$ companion solution with the same symmetry, unlike (\ref{SO6SolIIARescaled}), (\ref{SO7SolIIA}). 

We conclude by noting that the formulae (\ref{KKSU3sectorinIIA}), (\ref{KKSU3FieldStrengths}) can be also used to uplift the two non-supersymmetric critical points with SU(3) symmetry, using the numerical values given in table 3 of \cite{Guarino:2015qaa}. The resulting massive IIA configurations are qualitatively similar, though $\cN=0$, to the solution (\ref{SU3IIASolutionRescaled}). By the spectrum analysis of \cite{Guarino:2015qaa}, these two solutions can be declared to be stable under perturbations contained in $D=4$ $\cN=8$ dyonic ISO(7) supergravity.

\subsection{Flux quantisation and free energies} \label{subsec:FluxQuant}

As we will now show, the classical solutions of the previous section survive flux quantisation. Thus, the perturbatively stable ones extend to well defined massive type IIA string theory backgrounds. We will restrict our attention to the topologically $S^6$ solutions, and will not consider the quantisation of the more general classes of solutions that we discussed. Flux quantisation for the Behrndt-Cvetic \cite{Behrndt:2004km} solution (\ref{G2N=1SolIIARescaled}) has already been studied in \cite{Cassani:2009ck}. See also \cite{Tomasiello:2010zz} for general considerations about flux quantisation of AdS$_4$ solutions in massive type IIA.

The classical solutions presented in sections \ref{subsec:susysols} and \ref{subsec:nonsusysols} depend on the gauge coupling constants $g$ and $m$ of the $D=4$ supergravity. We have given this dependence in terms of two other constants, $L$ and ${\phi_0}$,
\begin{equation} \label{eq:Lgsintermsofgandm}
L^2 \propto g^{-\frac{25}{12}} \,  m^{\frac{1}{12}}  \; , \qquad 
e^{\phi_0} \propto g^{\frac{5}{6}} \,  m^{-\frac{5}{6}} \; , 
\end{equation}
more amenable to $D=10$ interpretation: $L$ sets the scale of the AdS$_4$ and $S^6$ radii and $e^{\phi_0}$ is proportional to the string couplig constant $g_s$. The two proportionality coefficients in (\ref{eq:Lgsintermsofgandm}) are dimensionless, order one and solution-dependent --we have given them below each solution. Classically, $L^2$ and $e^{\phi_0}$ can be rescaled to one without loss of generality using global symmetries of type IIA supergravity, see {\it e.g.}~(A.7), (A.9) of \cite{Guarino:2015vca}. At the quantum level, however, $L$ and $e^{\phi_0}$ become fixed in terms of the integral Page charges $k$ and $N$ associated to $\hat F_\0$ and $\hat F_\6$,
\begin{eqnarray} \label{quantFlux}
&& k = 2\pi \ell_s  \,  \hat F_\0 \equiv 2\pi \ell_s  \,  m  \; , \nonumber \\[5pt]
&& N = -\frac{1}{(2\pi \ell_s )^5 }  \int_{S^6}  e^{\frac12 \hat \phi} \   \hat{*} \hat F_\4 + \hat B_\2 \wedge d\hat A_\3 + \frac16 m \,  \hat B_\2 \wedge \hat B_\2 \wedge \hat B_\2  \; , 
\end{eqnarray}
where $\ell_s = \sqrt{\alpha^\prime}$ is the string length. Note that $d$ of the integrand in the $\hat F_\6$ quantisation condition gives the left-hand-side of the $\hat F_\4$ equation of motion, see (A.5) of \cite{Guarino:2015vca}. From a ten-dimensional perspective, the integers $k$ and $N$ respectively count the quanta of Romans mass and the number of D2--branes. 

Working out the integral in (\ref{quantFlux}) for each individual solution, we can solve for $L$ and $e^{\phi_0}$ in terms of $k$ and $N$ on a case-by-case basis as
\begin{eqnarray} \label{eq:Lphi}
L \propto \, N^{\frac{5}{24}} \, k^{\frac{1}{24}} \, \ell_s \; , \qquad 
e^{\phi_0} \propto \,  N^{-\frac{1}{6}}  \, k^{-\frac{5}{6}} \; ,
\end{eqnarray}
where the proportionality coefficients are again dimensionless, solution-dependent and order one. More concretely:
\begin{equation} \label{eq:LphiSols}
\textrm{
\begin{tabular}{lllll}
$\cN=2 \ , \; \textrm{U}(3) $  & : &  
$
L =  2^{-\frac{7}{48}} \, 3^{-\frac{7}{24}} \, \pi^{\frac{3}{8}}  \,  N^{\frac{5}{24}} \, k^{\frac{1}{24}} \, \ell_s
$  & , & 
$
e^{\phi_0} = 2^{\frac{11}{12}}  \,3^{-\frac{1}{6}}  \,\pi^{\frac{1}{2}}  \,  N^{-\frac{1}{6}}  \, k^{-\frac{5}{6}}
$  , \\[5pt]
$\cN=1 \ , \; \textrm{G}_2$  & : &  
$
L =  2^{-\frac{13}{8}} \, 3^{-\frac{5}{48}} \, 5^{\frac{11}{16}} \, \pi^{\frac{3}{8}}  \,  N^{\frac{5}{24}} \, k^{\frac{1}{24}} \, \ell_s
$  & , & 
$
e^{\phi_0} = 2^{\frac{1}{2}}  \,3^{\frac{1}{12}}  \,5^{\frac{1}{4}}  \,\pi^{\frac{1}{2}}  \,  N^{-\frac{1}{6}}  \, k^{-\frac{5}{6}}
$ , \\[5pt]
$\cN=1 \ , \; \textrm{SU}(3)$  & : &  
$
L = 2^{-\frac{19}{12}} \, 3^{-\frac{7}{24}}  \, 5^{\frac{11}{16}} \, \pi^{\frac{3}{8}} \, N^{\frac{5}{24}} \, k^{\frac{1}{24}} \, \ell_s 
$  & , & 
$
e^{\phi_0} = 2^{\frac{5}{3}} \, 3^{-\frac{1}{6}}  \, 5^{\frac{1}{4}} \, \pi^{\frac{1}{2}} \,  N^{-\frac{1}{6}}  \, k^{-\frac{5}{6}} 
$  , \\[5pt]
$\cN=0 \ , \; \textrm{SO}(7)_+$  & : &  
$
L = 2^{\frac{2}{3}} \, 3^{\frac{5}{24}}  \, 5^{-\frac{25}{48}} \, \pi^{\frac{3}{8}} \, N^{\frac{5}{24}} \, k^{\frac{1}{24}} \, \ell_s 
$  & , & 
$
e^{\phi_0} = 2^{\frac{2}{3}} \, 3^{-\frac{1}{6}}  \, 5^{\frac{5}{12}} \, \pi^{\frac{1}{2}} \,  N^{-\frac{1}{6}}  \, k^{-\frac{5}{6}} 
$  , \\[5pt]
$\cN=0 \ , \; \textrm{SO}(6)_+$  & : &  
$
L = 2^{-1} \, 3^{\frac{5}{24}}  \,  \pi^{\frac{3}{8}} \, N^{\frac{5}{24}} \, k^{\frac{1}{24}} \, \ell_s 
$  & , & 
$
e^{\phi_0} = 2^{2} \, 3^{-\frac{1}{6}}  \, \pi^{\frac{1}{2}} \,  N^{-\frac{1}{6}}  \, k^{-\frac{5}{6}} 
$  , \\[5pt]
$\cN=0 \ , \; \textrm{G}_2$  & : &  
$
L =  2^{-\frac{3}{4}}  \, 3^{\frac{19}{48}}  \,  \pi^{\frac{3}{8}} \, N^{\frac{5}{24}} \, k^{\frac{1}{24}} \, \ell_s 
$  & , & 
$
e^{\phi_0} =  2 \cdot 3^{\frac{1}{12}}  \, \pi^{\frac{1}{2}} \,  N^{-\frac{1}{6}}  \, k^{-\frac{5}{6}} 
$  . \\[5pt]
\end{tabular}
}
\end{equation}
In our topologically $S^6$ solutions, the only flux quantisation conditions that can be imposed are (\ref{quantFlux}) and thus the classical constants $L$ and $e^{\phi_0}$ in each solution become univocally fixed in terms of the quantum fluxes through (\ref{eq:LphiSols}). Although we have carried out this calculation for all solutions, stable or unstable, the flux quantisation analysis is only sensible in the former case. Thus, the perturbatively stable solutions (in particular, the supersymmetric solutions) can be promoted to solutions at the quantum level. They provide good massive type IIA string theory backgrounds.

This case-by-case analysis can also be used to solve for the classical constants $g$ and $m$ of the $D=4$ supergravity in terms of the quantum numbers $N$ and $k$. We find\footnote{ Perhaps a more suggestive rewrite for $g$ is $g^5 = 5 \, v(S^6) \,  (2\pi \ell_s)^{-5} \, N^{-1}$, with $v(S^6) = 16 \pi^3 /15 $ the volume of the unit radius round six-sphere.}
\begin{equation} \label{gmInTermsOfkN}
g = 2^{\frac{4}{5}} \, 3^{-\frac{1}{5}} \, \pi^{\frac{3}{5}} \, (2\pi \ell_s)^{-1} \,  N^{-\frac{1}{5}} \; , \qquad 
m = (2\pi \ell_s)^{-1}  \,   k \; , 
\end{equation}
with the same proportionality coefficients for all solutions, as it of course must be given that $g$, $m$ and $N$, $k$ characterise the theory, not only the individual solutions. The second relation in (\ref{gmInTermsOfkN}) is a mere rearrangement of the first equation in (\ref{quantFlux}). However, the fact that we do get the same expression for $g$ in terms of $N$ for all solutions provides a non-trivial crosscheck on our calculations. One might have naively thought that the dimensionless electric and magnetic couplings, $g^\prime =  2\pi \ell_s \, g$ and $ m^\prime \equiv k = 2\pi \ell_s \, m$, of the $D=4$ supergravity should obey the Dirac quantisation condition, $g^\prime m^\prime =2\pi n $, for some integer $n$. It is evident from (\ref{gmInTermsOfkN}) that this is not true. In retrospect, the IIA embedding of the $D=4$ supergravity makes it clear why this relation will not hold: Dirac quantisation conditions must hold for  Poincar\'e-dual Ramond-Ramond fluxes $\hat F_{(p+2)}$ and $\hat F_{(8-p)}$ \cite{Polchinski:1995mt}; but $\hat F_\0$ and $\hat F_\6$, to which $m$ and $g$ are respectively related, are not dual to each other. A purely four-dimensional argument can be also put forward: the $D=4$ theory can be always formulated in a symplectic duality frame where all charges are electric and there are no magnetic charges.

It was argued in \cite{Aharony:2010af} that, in contrast to the celebrated massless case, massive type IIA string theory cannot reach a strong coupling regime. This phenomenon was illustrated with some of the (analytically or numerically) known $\textrm{AdS}_4 \times \mathbb{CP}^3$ backgrounds. It was found that the string frame curvature radius in string units, $L_{\textrm{string}}$, and the string coupling $g_s \sim e^{\phi_0}$ both stay bounded for all values of the fluxes on those backgrounds. The string coupling $g_s$ was shown to initially increase with $N$, as in the massless IIA $\cN=6$ solution (corresponding to the IIA reduction of the $D=11$ Freund-Rubin solution \cite{Freund:1980xh}), but then reach a second phase where $g_s$ decreases as  $L_{\textrm{string}}$ becomes large for $N$ large compared to $k$:  $L_{\textrm{string}} \propto \, N^{\frac{1}{6}} \, k^{-\frac{1}{6}} \, \ell_s$ and $e^{\phi_0} \propto \,  N^{-\frac{1}{6}}  \, k^{-\frac{5}{6}}$. All our $\textrm{AdS}_4 \times S^6$ solutions behave as in this second phase for all values of the fluxes, as can be seen by translating the individual values of the Einstein frame $L$ in (\ref{eq:LphiSols}) into the string frame via $L_{\textrm{string}} = L \, e^{\frac{1}{4} \phi_0}$. Note also that $k$ here corresponds to $n_0$ in  \cite{Aharony:2010af}. 

We conclude with the calculation of the gravitational free energy of these solutions. The gravitational free energy $F$ of a unit radius AdS$_4$ solution is given by $F= \pi/(2 G_4)$ \cite{Emparan:1999pm}, where $G_4$ is the effective four-dimensional Newton's constant. Plugging the solutions into the ten-dimensional action (see (A.1) of \cite{Guarino:2015vca} for our conventions), we obtain
\begin{eqnarray} \label{FreeEnergy1}
F= \frac{16 \pi^3}{(2\pi \ell_s )^8 }  \int_{S^6} e^{8A} \,  \textrm{vol}_6 \; , 
\end{eqnarray}
where $e^{2A}$ is the warp factor (\ref{eq:WarpFactorandFR}) and $\textrm{vol}_6$ is the volume form corresponding to the internal metric $ds_6^2$ in (\ref{GenericFormMetric}). We have evaluated this integral for each solution, and have traded the resulting $L^8$ dependence for the quantum numbers $N$ and $k$ via (\ref{eq:LphiSols}). The results appear in table \ref{Table:SU3Points} of the introduction. For completeness, the table also shows the free energy of the $\cN=2$ solution, which was already given in \cite{Guarino:2015jca}. These free energies turn out to be inversely proportional to the four-dimensional cosmological constant at the corresponding $D=4$ critical point. Indeed, for any two free energies $F_1$ and $F_2$ in table \ref{Table:SU3Points}, and any $D=4$ cosmological constants $V_1$, $V_2$ in table 3 of \cite{Guarino:2015qaa}, it is straightforward to verify that 
\begin{eqnarray}
\frac{F_1}{F_2} = \frac{V_2}{V_1} \; .
\end{eqnarray}

The AdS$_4$ solutions contained in this paper should be dual to Chern-Simons-matter theories with a single gauge group SU$(N)$ at level $k$ of the type considered in \cite{Schwarz:2004yj}. A match of gravitational and field theory free energies was given for the $\cN=2$ solution in \cite{Guarino:2015jca}. It would be very interesting to understand better the holography of the $\cN=1$ solutions.

\section*{Acknowledgements}

Adolfo Guarino and Daniel Jafferis are warmly acknowledged for collaboration in related projects and fruitful discussions. Conversations with Alessandro Tomasiello were also very useful. This work was supported by the Marie Curie fellowship PIOF-GA-2012-328798, managed from the CPHT of \'Ecole Polytechnique, and partially by the Fundamental Laws Initiative at Harvard. This paper was completed during the {\it Stringy Geometry} workshop held at the MITP of Mainz in September 2015. Hospitality and partial support from MITP are gratefully acknowledged.

%%%%%%%%%%%%%%%%%
%%%%%%%%%%%%%%%%%
\appendix
%%%%%%%%%%%%%%%%%
%%%%%%%%%%%%%%%%%

\newpage

%%%%%%%%%%%%%%%%%
%%%%%%%%%%%%%%%%%

\section{\mbox{Foliation of $S^6$ with Sasaki-Einstein leaves}} \label{subset:SE5S6}
%%%%%%%%%%%%%%%%%
%%%%%%%%%%%%%%%%%

Here we describe how the natural embedding of $S^6$ in $\mathbb{R}^7$ determines the foliation of the former with $S^5$ leaves, the latter equipped with their natural Sasaki-Einstein structure. Let $\mu^I$, $I=1, \ldots, 7$, parametrise $S^6$ as a the locus $\delta_{IJ} \mu^I \mu^J = 1$ in $\mathbb{R}^7$, and let $y^m$, $m=1, \ldots, 6$,  be the $S^6$ angles. It is convenient to split the index $I = (i,7)$, $i=1, \ldots, 6$, and the $\mu^I$ as
\begin{eqnarray} \label{eq:SplitS6intoS5}
\mu^i = \sin \alpha \, \tilde \mu^i \; , \; i=1, \ldots, 6 \; , \qquad 
\mu^7 = \cos \alpha \; , 
\end{eqnarray}
where
\begin{equation} \label{anglealpha}
0 \leq \alpha \leq \pi
\end{equation}
is one of the $S^6$ angles $y^m$, and $\tilde \mu^i$ characterise the $S^5$ within $S^6$ as the locus $\delta_{ij} \tilde{\mu}^i \tilde{\mu}^j = 1$ in $\mathbb{R}^6 \subset \mathbb{R}^7$. In terms of (\ref{eq:SplitS6intoS5}), the round metric on $S^6$ with radius $g$, given in {\it e.g.}~(E.2) of \cite{Guarino:2015vca} with $n=6$, acquires the familiar sine-cone form
\begin{eqnarray} \label{eq:sineconeS6}
d\mathring{s}^2(S^6)  = g^{-2} \, \Big(  d\alpha^2 + \sin^2 \alpha \;d\tilde{s}^2(S^5)   \Big)  \; , 
\end{eqnarray}
where $ d\tilde{s}^2(S^5)$ is the round metric (E.2) of \cite{Guarino:2015vca} with $n=5$ and $g=1$. Denoting by $\tilde m = 1 , \ldots , 5$ the directions along the $S^5$, the Killing vectors $K^{m IJ} = 2 g^{-2} \mathring{g}^{mn} \mu^{[I} \partial_n \mu^{J]}$ of the round metric (\ref{eq:sineconeS6}) are
\begin{eqnarray} \label{KillingFoliation}
K^\alpha_{ij} = 0 \; , \quad
K^{\tilde{m}}_{ij} = \tilde{K}^{\tilde{m}}_{ij} \; , \quad 
K^\alpha_{i7} = -\tilde \mu_i \; , \quad
K^{\tilde m}_{i7} = - \cot \alpha \,  \tilde{g}^{mn} \partial_n \tilde{\mu}_i \; , 
\end{eqnarray}
and the tensors $K_{mn}^{IJ} = 4g^{-2} \partial_{[m} \mu^J \partial_{n]} \mu^J$ are
\begin{eqnarray} \label{KillingDerFoliation}
&& K^{ij}_{\alpha \tilde m} = 4 g^{-2} \sin \alpha \cos \alpha \, \tilde{\mu}^{[i} \partial_{\tilde m} \tilde \mu^{j]} \; , \quad 
K^{ij}_{\tilde m \tilde n} = g^{-2}  \sin^2\alpha \, \tilde{K}^{ij}_{\tilde m \tilde n} \; , \nonumber \\
&& K^{i7}_{\alpha \tilde m} = 2 g^{-2} \sin^2 \alpha \, \partial_{\tilde m} \tilde{\mu}^i \; ,  \qquad \qquad \;
K_{\tilde m \tilde n}^{i7} = 0 \; . 
\end{eqnarray}
In the above expressions, $\partial_m \mu^I$ means derivative of $\mu^I = \mu^I(y^m)$ with respect to $y^m$.

Let us now consider the canonical Sasaki-Einstein structure on $S^5$. Recall, more generally, that a Sasaki-Einstein structure on a five-manifold $M_5$ is an SU(2)--structure, therefore determined by a real $(1,1)$-form $\bm{J}$, a complex decomposable $(2,0)$-form $\bm{\Omega}$, and a real one-form $\bm{\eta}$. These forms are subject to the  following algebraic
\begin{eqnarray} \label{SE5str}
\iota_{\partial_\psi} \bm{J} =\iota_{\partial_\psi} \bm{\Omega} = 0   \; , \qquad 
\bm{\Omega} \wedge \bm{\bar \Omega} = 2 \, \bm{J} \wedge \bm{J}  \neq 0 \; , \qquad 
\bm{J} \wedge \bm{\Omega} = 0  \ ,
\end{eqnarray}
and differential relations 
\begin{eqnarray} \label{SE5strDif}
d \bm{\eta} = 2 \bm{J} \; , \qquad 
d \bm{J} =  0 \; , \qquad 
d \bm{\Omega} = 3 i\,  \bm{\Omega} \wedge \bm{\eta}  \ . 
\end{eqnarray}
In (\ref{SE5str}), we have introduced a U(1) angle $\psi$ so that  the vector $\partial_\psi$ is dual to $\bm{\eta}$. We can locally write $\bm{\eta} = d\psi + \sigma$, where $\sigma$ is a one-form potential for $\bm{J}$, with $d \sigma = 2 \bm{J}$. This SU(2)--structure specifies an Einstein metric $ds^2(M_5)$, normalised so that the Ricci tensor is $4$ times the metric. Locally, this metric can be written as
\begin{eqnarray} \label{SE5metric}
ds^2(M_5) = ds^2(\textrm{KE}_4) + \bm{\eta}^2 \; ,
\end{eqnarray}
where $ds^2(\textrm{KE}_4)$ is a metric defined, at least locally, on a positively curved K\"ahler-Einstein manifold or orbifold, normalised so that the Ricci tensor is 6 times the metric. The vector $\partial_\psi$ is a Killing vector of (\ref{SE5metric}). The cone $C(M_5) = \mathbb{R}^+ \times M_5$ over $M_5$ has SU(3) holonomy. The $(1,1)$--form $J$ and $(3,0)$--form $\Omega$ that the latter condition defines on $C(M_5)$ can be written as
\begin{eqnarray} \label{SU3SE5}
J = r^2 \bm{J} + r dr \wedge \bm{\eta}   \; , \qquad 
\Omega = r^2 ( dr +i r  \bm{\eta} ) \wedge \bm{\Omega}  \; , 
\end{eqnarray}
where $r$ is a coordinate on the $ \mathbb{R}^+$ factor. The closure of $J$ and $\Omega$, required by SU(3)--holonomy, is equivalent to the Sasaki-Einstein torsion conditions (\ref{SE5strDif}). 

For the specific case of $S^5$, we have that $\textrm{KE}_4 = \mathbb{CP}^2$ and $C(S^5) = \mathbb{R}^6$. Following a similar strategy described in appendix E of \cite{Guarino:2015vca}, the Sasaki-Einstein forms $\bm{\eta}$, $\bm{J}$,  $\bm{ \Omega}$ on $S^5$ can be written in terms of the embedding coordinates $\tilde{\mu}^i$ and the components $J_{ij}$, $\Omega_{ijk}$ of the SU(3)--holonomy forms on $\mathbb{R}^6$ as 
\begin{eqnarray} \label{SE5intermsofmu}
\bm{\eta} = J_{ij} \,  \tilde{\mu}^i  d\tilde{\mu}^j  \; , \qquad 
\bm{J} = \tfrac12 \, J_{ij} \,  d\tilde{\mu}^i \wedge d\tilde{\mu}^j  \; , \qquad 
 \bm{ \Omega}   =\tfrac12 \, \Omega_{ijk} \, \tilde{\mu}^i \, d \tilde{\mu}^j \wedge d\tilde{\mu}^k  \; . 
\end{eqnarray}
These expressions are crucial to bring the consistent truncation formulae of type IIA down to the SU(3)-invariant sector of ISO(7) supergravity to the form (\ref{KKSU3sectorinIIA}). 

We conclude with the relation of the homogeneous nearly-K\"ahler structure ${\cal J}$, $\Upomega$ on $S^6$ and the canonical Sasaki-Einstein structure of its foliating $S^5$. Introducing a frame $e^I$, $I=1, \ldots, 7$ on $\mathbb{R}^7$, and splitting $\mathbb{R}^7 = \mathbb{R}^6 \times \mathbb{R}$, the associative and coassociative forms $\psi$, $\tilde{\psi}$ on $\mathbb{R}^7$ can be written in terms of the SU(3)--holonomy forms $J$, $\Omega$ on the $\mathbb{R}^6$ factor as 
\begin{eqnarray} \label{eq:G2intermsofSU3}
\psi = J \wedge e^7 +  \textrm{Re} \,  \Omega \; , \qquad 
\tilde{\psi} = \tfrac12 J \wedge J + \textrm{Im} \, \Omega \wedge e^7 \; .
\end{eqnarray}
From here, the following non-vanishing components can be read off:
\begin{eqnarray} \label{eq:G2intermsofSU3Comps}
\psi_{ijk} = (\textrm{Re} \,  \Omega)_{ijk} \; , \quad 
\psi_{ij7} = J_{ij} \; ,  \quad
\tilde{\psi}_{ijk\ell} = 3 J_{[ij} J_{k \ell]} \; , \quad 
\tilde{\psi}_{ijk7} = (\textrm{Im} \,  \Omega)_{ijk} \; . \quad 
\end{eqnarray}
Finally, introducing (\ref{eq:G2intermsofSU3Comps}) and (\ref{eq:SplitS6intoS5}) into 
\begin{eqnarray} \label{JOmegaintermsofmu}
{\cal J} = \tfrac12 \, \psi_{IJK} \,  \mu^I d\mu^J \wedge d\mu^K \; , \quad 
  \Upomega  = \tfrac16 \left( \psi_{JKL} -i \, \tilde{\psi}_{IJKL} \, \mu^I \right)  d \mu^J \wedge d\mu^K \wedge d\mu^L 
\end{eqnarray}
(see appendix E of \cite{Guarino:2015vca}), the homogeneous nearly-K\"ahler forms on $S^6$ can be written in terms of the Sasaki-Einstein forms on the foliating $S^5$ as
\begin{eqnarray} \label{NKintermsof SE5Forms}
&& {\cal J} = \sin^2\alpha \cos \alpha \, \bm{J} +\sin^3 \alpha \, \textrm{Re} \, \bm{\Omega} + \sin \alpha \, d\alpha \wedge \bm{\eta} \; , \nonumber \\
&& \textrm{Re} \, \Upomega = -\sin^3 \alpha \, \bm{J} \wedge d\alpha  + \sin^2\alpha \cos \alpha \, \textrm{Re} \, \bm{\Omega }\wedge d\alpha -\sin^3\alpha \, \textrm{Im} \, \bm{\Omega } \wedge \bm{\eta} \; , \nonumber \\
&& \textrm{Im} \, \Upomega = -\sin^4 \alpha \, \bm{J} \wedge \bm{\eta}  + \sin^3\alpha \cos \alpha \, \textrm{Re} \, \bm{\Omega }\wedge \bm{\eta} +\sin^2\alpha \, \textrm{Im} \, \bm{\Omega } \wedge d\alpha  \; .
\end{eqnarray}
Using the Sasaki-Einstein conditions (\ref{SE5str}), (\ref{SE5strDif}) for $\bm{J}, \bm{\Omega} , \bm{\eta}$, the forms ${\cal J}, \Upomega$ in (\ref{NKintermsof SE5Forms}) can indeed be doublechecked to satisfy the nearly-K\"ahler relations 
\begin{eqnarray} \label{SU3str}
\Upomega \wedge \bar \Upomega = -\tfrac{4i}{3} {\cal J} \wedge {\cal J} \wedge {\cal J} \neq 0 \; , \quad {\cal J} \wedge \Upomega = 0  \ ,
\end{eqnarray}
and
\begin{eqnarray} \label{SU3strDif}
d {\cal J} = 3 \,  \textrm{Re} \,  \Upomega   \; , \quad d \, \textrm{Im} \, \Upomega =  -2 \,  {\cal J} \wedge {\cal J} \ . 
\end{eqnarray}
%
%

%%%%%%%%%%%%%%%
%%%%%%%%%%%%%%%
\section{SU(3)--invariant AdS$_4$ solutions of massive type IIA} \label{SU3InvAdS4Sols}
%%%%%%%%%%%%%%%
%%%%%%%%%%%%%%%

The solutions we have presented in the main text are of the local form\footnote{The warp factor and the Freund-Rubin constant were denoted in the main text by $e^{2A}$ and $U_{\textrm{rescaled}}$: see (\ref{eq:WarpFactorandFR}). In this appendix, these quantities are denoted by $e^X$ and $\mu_0$. The function $X$ should not cause any confusion with $X$ defined in the main text, in (\ref{eq:XY}). Also in this appendix, $e^{2A}$ is used to denote a metric function different than the warp factor. This should not cause any confusion either. Finally, note that the G$_2$-invariant solutions (\ref{G2N=1SolIIARescaled}), (\ref{G2N=0SolIIARescaled}) can also be brought to the local form (\ref{SU3IIAconfig}) through the identifications (\ref{NKintermsof SE5Forms}).}
{\setlength\arraycolsep{0pt}
\begin{eqnarray} \label{SU3IIAconfig}
&& d\hat{s}_{10}^2 = e^{2X(\alpha)} ds^2 (\textrm{AdS}_4) + e^{2A(\alpha)} d\alpha^2 + e^{2B(\alpha)} ds^2 (\textrm{KE}_4) + e^{2C(\alpha)} \bm{\eta}^2 \; , 
 \qquad \hat \phi = \phi(\alpha) \; ,  \nonumber \\[8pt]
&& \hat F_\4 = \mu_0 \textrm{vol}_4 + A_4(\alpha) \,  \textrm{vol} (\textrm{KE}_4) +B_4(\alpha) \, \bm{J} \wedge d\alpha \wedge \bm{\eta} 
\nonumber \\
&& \qquad \; + C_4 (\alpha) \, \textrm{Re} \, \bm{\Omega} \wedge d\alpha \wedge \bm{\eta}  + D_4 (\alpha) \, \textrm{Im} \, \bm{\Omega} \wedge d\alpha \wedge \bm{\eta} \; ,   \nonumber \\[8pt]
&& \hat H_\3 = B_3(\alpha) \, \bm{J} \wedge d\alpha + C_3 (\alpha) \, \textrm{Re} \, \bm{\Omega} \wedge d\alpha  + D_3 (\alpha) \, \textrm{Im} \, \bm{\Omega} \wedge d\alpha \nonumber \\
&& \qquad \; + E_3 (\alpha) \, \textrm{Re} \, \bm{\Omega} \wedge \bm{\eta}  + F_3 (\alpha) \, \textrm{Im} \, \bm{\Omega} \wedge  \bm{\eta} \; ,  \nonumber \\[8pt]
&& \hat F_\2 = A_2(\alpha)  \, \bm{J }  +B_2(\alpha) \, d\alpha \wedge \bm{\eta}  +C_2 (\alpha) \, \textrm{Re} \, \bm{\Omega} + D_2 (\alpha) \, \textrm{Im} \, \bm{\Omega} \; ,
\end{eqnarray}
}where AdS$_4$ is unit radius, $\mu_0$ is a constant, $X(\alpha)$, etc., are functions on the angle $\alpha$ and $\bm{J}$, $\bm{\Omega}$, $\bm{\eta}$ are the usual forms on a Sasaki-Einstein five-manifold $M_5$ (see appendix \ref{subset:SE5S6}) with metric $ ds^2 (\textrm{KE}_4)$ on the local K\"ahler-Einstein base normalised so that the Ricci tensor equals 6 times the metric. When $M_5 = S^5$ so that $\textrm{KE}_4 = \mathbb{CP}^2$, (\ref{SU3IIAconfig}) is the most general AdS$_4$ configuration of type IIA supergravity with SU(3) symmetry. In this appendix we work out the differential and algebraic equations that the functions $X(\alpha)$, etc., must obey for (\ref{SU3IIAconfig}) to solve the Bianchi identities and equations of motion of massive type IIA supergravity. We have used the equations in this appendix to verify that the solutions presented in the main text, that we obtained by using the uplifting formulae  (\ref{KKSU3sectorinIIA}), (\ref{KKSU3FieldStrengths}), do indeed solve the type IIA field equations.

In order to write the Einstein equation below, we find it convenient to introduce the local internal metric
\begin{equation} \label{Met6Conf}
ds_6^2 = e^{2 X (\alpha)} \Big(  e^{2A(\alpha)} d\alpha^2 + e^{2B(\alpha)} ds^2 (\textrm{KE}_4) + e^{2C(\alpha)} \bm{\eta}^2 \Big) \; , 
\end{equation}
which is conformal to the internal metric that appears in (\ref{SU3IIAconfig}). The Ricci tensor of the metric (\ref{Met6Conf}) with curved indices reads
\begin{eqnarray} \label{RicciConformalmet6}
R_{\alpha \alpha} & = & - \big(  4X -A +4B +C  \big)^{\prime\prime} - e^{-4X+A-4B-C} \big(  e^{4X-A+4B+C} (X+A)^\prime  \big)^\prime \nonumber \\
&& + (X+A)^\prime  (9X-A+8B+2C)^\prime - 4 \big((X+B)^\prime \big)^2-  \big((X+C)^\prime \big)^2 \; ,
\nonumber \\[8pt]
R_{mn} & = & -\Big[ e^{- 4X -A -2B -C } \big(  e^{4X-A+4B+C} (X+B)^\prime  \big)^\prime +2 \,  e^{-2B +2C } -6 \Big] \,  g_{mn}  \; ,  
\nonumber \\[8pt]
R_{\psi \psi} & = & - e^{- 4X -A -4B +C } \big(  e^{4X-A+4B+C} (X+C)^\prime  \big)^\prime +4 \,  e^{-4B +4C } \; , 
\end{eqnarray}
with $g_{mn}$ the components of $ds^2 (\textrm{KE}_4)$, and $R_{\alpha m}  = R_{\alpha \psi} = R_{m \psi} =0$. Here and in the following, we drop the explicit $\alpha$ dependence and denote with a prime the derivative with respect to it.

The IIA Bianchi identities (see (A.3) of \cite{Guarino:2015vca}) impose the restrictions
{\setlength\arraycolsep{0pt}
\begin{eqnarray} \label{SU3IIABianchis}
&& A_4^\prime -4B_4 -2A_2 B_3 -2C_2 C_3 -3D_2D_3 =0 \;, \nonumber \\
&& C_2 E_3 + D_2 F_3 = 0 \; ,  \nonumber \\
&& E_3^\prime -3 D_3 = 0 \; ,  \nonumber \\
&& F_3^\prime +3 C_3 = 0 \; ,  \nonumber \\
&& A_2^\prime -2B_2 -m B_3 = 0 \; ,  \nonumber \\
&& C_2^\prime -m C_3 = 0 \; ,  \nonumber \\
&& D_2^\prime -m D_3 = 0 \; ,  \nonumber \\
&& 3D_2 -m E_3 = 0 \; ,  \nonumber \\
&& 3C_2 +m F_3 = 0 \; .
\end{eqnarray}
}Turning now to the equations of motion (see (A.5) of \cite{Guarino:2015vca}), we find that the $\hat F_\4$ equation of motion gives
{\setlength\arraycolsep{0pt}
\begin{eqnarray} \label{SU3IIAF4eom}
&&  \big( e^{ \frac12 \phi +4X-A-C }  B_4 \big)^\prime - 2  \, e^{ \frac12 \phi +4X + A -4B+C }  A_4 + \mu_0 B_3 =0 \;, \nonumber \\
&&  \big( e^{ \frac12 \phi +4X-A-C }  C_4 \big)^\prime + \mu_0 C_3 =0 \;, \nonumber \\
&&  \big( e^{ \frac12 \phi +4X-A-C }  D_4 \big)^\prime + \mu_0 D_3 =0 \;, \nonumber \\
&&  3  \, e^{ \frac12 \phi +4X - A -C }  D_4 + \mu_0 E_3 =0 \;, \nonumber \\
&&  3  \, e^{ \frac12 \phi +4X - A -C }  C_4 - \mu_0 F_3 =0 \;,
\end{eqnarray}
}the $\hat H_\3$ equation of motion gives
{\setlength\arraycolsep{0pt}
\begin{eqnarray} \label{SU3IIAH3eom}
&&  \big( e^{ - \phi +4X-A+C }  B_3 \big)^\prime -  e^{ \frac12 \phi +4X + A -4B+C }  A_2 A_4 -  e^{ \frac12 \phi +4X - A -C }  B_2 B_4  \nonumber \\
&& \qquad \qquad  - m  \, e^{ \frac32 \phi +4X + A +C } A_2  - \mu_0 B_4 =0 \;, \nonumber \\[10pt]
&&  \big( e^{ - \phi +4X-A+C }  C_3 \big)^\prime -  e^{ \frac12 \phi +4X + A -4B+C }  C_2 A_4 -  e^{ \frac12 \phi +4X - A -C }  B_2 C_4 + 3\, e^{ -\phi +4X + A -C }  F_3   \nonumber \\
&& \qquad \qquad  - m  \, e^{ \frac32 \phi +4X + A +C } C_2  - \mu_0 C_4 =0 \;, \nonumber \\[10pt]
&&  \big( e^{ - \phi +4X-A+C }  D_3 \big)^\prime -  e^{ \frac12 \phi +4X + A -4B+C }  D_2 A_4 -  e^{ \frac12 \phi +4X - A -C }  B_2 D_4 - 3\, e^{ -\phi +4X + A -C }  E_3   \nonumber \\
&& \qquad \qquad  - m  \, e^{ \frac32 \phi +4X + A +C } D_2  - \mu_0 D_4 =0 \;, \nonumber \\[10pt]
&& e^{ - \phi +4X-A+C }  B_3 -\tfrac12 \,  e^{ \frac12 \phi +4X - A -C } \big(  A_2 B_4 + C_2 C_4  + D_2 D_4  \big) \nonumber \\
&& \qquad \qquad  -\tfrac14 m  \, e^{ \frac32 \phi +4X - A +4B -C } B_2  -\tfrac14  \mu_0 A_4 =0 \; , 
\end{eqnarray}
}the $\hat F_\2$ equation of motion gives
{\setlength\arraycolsep{0pt}
\begin{eqnarray} \label{SU3IIAF2eom}
&& \big( e^{ \frac32 \phi +4X-A+4B - C }  B_2 )^\prime -4 \,  e^{ \frac32 \phi +4X + A +C } A_2   +2  \,  e^{ \frac12 \phi +4X - A -C } \big(  B_3 B_4 + C_3 C_4  + D_3 D_4  \big) = 0 \; ,   \nonumber \\[5pt]
&&  E_3 C_4 + F_3 D_4 = 0 \; , 
\end{eqnarray}
}and the dilaton equation of motion gives

{\setlength\arraycolsep{0pt}
\begin{eqnarray} \label{SU3IIADilaton}
&& \big( e^{ 4X-A+4B + C }  \phi^\prime )^\prime -\tfrac32 \,  e^{ \frac32 \phi +4X + A +C } \big( A_2^2  + C_2^2 + D_2^2 \big)  - \tfrac34 e^{ \frac32 \phi +4X - A +4B -C } B_2^2  \nonumber \\
&& \qquad \qquad  + e^{ - \phi +4X - A +C } \big( B_3^2  + C_3^2 + D_3^2 \big) + e^{ - \phi +4X + A -C } \big( E_3^2  + F_3^2 \big)   \nonumber \\
&&\qquad \qquad  -\tfrac14   \,  e^{ \frac12 \phi +4X + A -4B +C } A_4^2   -\tfrac12   \,  e^{ \frac12 \phi +4X - A -C } \big( B_4^2  + C_4^2 + D_4^2 \big) \nonumber \\
&&  \qquad \qquad -\tfrac54   \, m^2 \,   e^{ \frac52 \phi +4X + A +4B +C }  +\tfrac14   \, \mu_0^2 \,   e^{ \frac12 \phi -4X + A +4B +C } = 0 \; . 
\end{eqnarray}
}Finally, the external components of the $D=10$ Einstein equation produce an equation for the warp factor,
{\setlength\arraycolsep{0pt}
\begin{eqnarray} \label{SU3IIAEinsteinExt}
&& \big( e^{ 4X-A+4B + C }  X^\prime )^\prime -\tfrac18 \,  e^{ \frac32 \phi +4X + A +C } \big( A_2^2  + C_2^2 + D_2^2 \big)  - \tfrac{1}{16} e^{ \frac32 \phi +4X - A +4B -C } B_2^2  \nonumber \\
&& \qquad \qquad   -\tfrac14\, e^{ - \phi +4X - A +C } \big( B_3^2  + C_3^2 + D_3^2 \big)    -\tfrac14\,  e^{ - \phi +4X + A -C } \big( E_3^2  + F_3^2 \big)  \nonumber \\
&&\qquad \qquad  -\tfrac{3}{16}   \,  e^{ \frac12 \phi +4X + A -4B +C } A_4^2   -\tfrac{3}{8}    \,  e^{ \frac12 \phi +4X - A -C } \big( B_4^2  + C_4^2 + D_4^2 \big) \nonumber  \\
&&  \qquad \qquad +\tfrac{1}{16}   \, m^2 \,   e^{ \frac52 \phi +4X + A +4B +C }  -\tfrac{5}{16}   \, \mu_0^2 \,   e^{ \frac12 \phi -4X + A +4B +C } + 3\, e^{ 2X + A +4B +C } = 0   ,  \;
\end{eqnarray}
}and the internal components give the equations
\begin{equation} \label{SU3IIAEinsteinInt}
R_{\alpha \alpha} = T_{\alpha \alpha}  \; , \qquad 
R_{mn} = T_{mn}  \; , \qquad 
R_{\psi \psi} = T_{\psi \psi}  \; , \qquad
C_3E_3 +D_3 F_3 = 0 \; , 
\end{equation}
where the left-hand-sides have been given in (\ref{RicciConformalmet6}), and the right-hand-sides are
{\setlength\arraycolsep{2pt}
\begin{eqnarray} \label{SU3IIARHSEinsteinInt}
T_{\alpha \alpha} & \equiv & \tfrac12 ( \phi^\prime  )^2 +  8 \,  ( X^\prime  )^2   -\tfrac14   \,  e^{ \frac32 \phi +2A-4B } \big( A_2^2  + C_2^2 + D_2^2 \big)  + \tfrac{3}{8} e^{ \frac32 \phi -2C } B_2^2  \nonumber \\
&&  +\tfrac12\, e^{ - \phi -4B} \big( B_3^2  + C_3^2 + D_3^2 \big)    -\tfrac12\,  e^{ - \phi +2A -4B -2C } \big( E_3^2  + F_3^2 \big)  \nonumber \\
&&  -\tfrac{3}{8}   \,  e^{ \frac12 \phi +2 A -8B } A_4^2   +\tfrac{1}{4}    \,  e^{ \frac12 \phi -4B -2C } \big( B_4^2  + C_4^2 + D_4^2 \big) \nonumber  \\
&&  +\tfrac{1}{8}   \, m^2 \,   e^{ \frac52 \phi + 2A }  -\tfrac{1}{8}   \, \mu_0^2 \,   e^{ \frac12 \phi -8X + 2A } + 3\, e^{ -2X +2 A }   \; ,  \nonumber \\[10pt]
T_{mn} & \equiv & \Big[ \tfrac14 e^{ \frac32 \phi -2B } \big( A_2^2  + C_2^2 + D_2^2 \big)  - \tfrac{1}{8} e^{ \frac32 \phi -2A+2B -2C } B_2^2  \nonumber \\
&&  +\tfrac{1}{8}   \,  e^{ \frac12 \phi -6B } A_4^2   -\tfrac{1}{4}    \,  e^{ \frac12 \phi -2A-2B -2C } \big( B_4^2  + C_4^2 + D_4^2 \big) \nonumber  \\
&&  +\tfrac{1}{8}   \, m^2 \,   e^{ \frac52 \phi + 2B }  -\tfrac{1}{8}   \, \mu_0^2 \,   e^{ \frac12 \phi -8X + 2B } + 3\, e^{ -2X +2 B } \Big] \,  g_{mn}    \; ,   \nonumber \\[10pt]
T_{\psi \psi} & \equiv &   -\tfrac14   \,  e^{ \frac32 \phi -4B +2C } \big( A_2^2  + C_2^2 + D_2^2 \big)  + \tfrac{3}{8} e^{ \frac32 \phi -2A } B_2^2  \nonumber \\
&&  -\tfrac12\, e^{ - \phi -2A-4B+2C} \big( B_3^2  + C_3^2 + D_3^2 \big)    +\tfrac12\,  e^{ - \phi -4B  } \big( E_3^2  + F_3^2 \big)  \nonumber \\
&&  -\tfrac{3}{8}   \,  e^{ \frac12 \phi -8B +2C } A_4^2   +\tfrac{1}{4}    \,  e^{ \frac12 \phi -2A -4B  } \big( B_4^2  + C_4^2 + D_4^2 \big) \nonumber  \\
&&  +\tfrac{1}{8}   \, m^2 \,   e^{ \frac52 \phi + 2C }  -\tfrac{1}{8}   \, \mu_0^2 \,   e^{ \frac12 \phi -8X + 2C } + 3\, e^{ -2X +2 C }   \; . 
\end{eqnarray}
}

We have explicitly verified that the solutions given in the main text solve all the equations (\ref{SU3IIABianchis})--(\ref{SU3IIAEinsteinInt}) and, hence, are indeed solutions of massive type IIA supergravity. It is apparent that without a solution generating technique like  (\ref{KKSU3sectorinIIA}), (\ref{KKSU3FieldStrengths}), it would have been extremely difficult to find analytic solutions in closed form to these equations.

\newpage

\bibliography{SU3solutions}
\end{document}